\documentclass[iop]{emulateapj}   %use this later on

\usepackage{comment}
\usepackage{hyperref}
\usepackage[flushleft]{threeparttable}

\usepackage{amsmath}
\usepackage{epstopdf}
\shorttitle{}
\shortauthors{}

\begin{document}

\title{Low star formation efficiency in typical galaxies at $z=5$--6}

\author{Riccardo Pavesi\altaffilmark{1}$^\dagger$, Dominik A. Riechers\altaffilmark{1,2}, Andreas L. Faisst\altaffilmark{3}, Gordon J. Stacey\altaffilmark{1}, Peter L. Capak\altaffilmark{3}
}

\affil{$^1$Department of Astronomy, Cornell University, Space Sciences
Building, Ithaca, NY 14853, USA\\
$^2$Max Planck Institute for Astronomy, K{\"o}nigstuhl 17, 69117 Heidelberg, Germany \\
$^3$Infrared Processing and Analysis Center, California Institute of Technology, Pasadena, CA 91125, USA
}
     \email{$^\dagger$rp462@cornell.edu}

\begin{abstract}
 Using the VLA and ALMA, we have obtained CO(2--1), [C{\sc ii}], [N{\sc ii}] line emission and multiple dust continuum measurements in a sample of ``normal" galaxies at $z=5-6$.
We report the highest redshift detection of low-$J$ CO emission from a  Lyman Break Galaxy, at $z\sim5.7$. The CO line luminosity implies a massive molecular gas reservoir of $(1.3\pm0.3)(\alpha_{\rm CO}/4.5\,M_\odot$ (K km s$^{-1}$ pc$^2)^{-1})\times10^{11}\,M_\odot$, suggesting low  star formation efficiency, with a gas depletion timescale of order $\sim$1 Gyr. This efficiency is much lower than traditionally observed in $z\gtrsim5$ starbursts, indicating that star forming conditions in Main Sequence galaxies at $z\sim6$ may be comparable to those of normal galaxies probed up to $z\sim3$ to-date, but with rising gas fractions across the entire redshift range.
We also obtain a deep CO upper limit for a Main Sequence galaxy at $z\sim5.3$ with $\sim3$ times lower SFR, perhaps implying a high $\alpha_{\rm CO}$ conversion factor, as typically found in low metallicity galaxies.
 For a sample including both CO targets, we also find faint [N{\sc ii}] 205$\,\mu$m emission relative to [C{\sc ii}] in all but the most IR-luminous ``normal" galaxies at $z=5-6$, implying more intense or harder radiation fields in the ionized gas relative to lower redshift. These radiation properties suggest that low metallicity may be common in typical $\sim$10$^{10}\,M_\odot$ galaxies at $z=5-6$.
While a fraction of Main Sequence star formation in the first billion years may take place in conditions not dissimilar to lower redshift, lower metallicity may affect the remainder of the population.

\end{abstract}

\section{Introduction}

Massive galaxies started forming during the epoch of reionization at $z>6$ and may have experienced their fastest growth toward the end of the first billion years of cosmic time ($z\sim4-6$), doubling their stellar mass content on time-scales of order a hundred million years (e.g., \citealt{Bouwens2015,Faisst16a}). While the high redshift universe offers the promise of strong new constraints to dark matter physics through early halo growth (e.g., \citealt{BuckleyDM}), they have, so far, been limited to coarse stellar mass-halo mass relationships which do not capture the variety in galaxy formation history hinted at by observations (e.g., \citealt{Behroozi2018,Moster2018,Tacchella2018}).
The details of such an early growth epoch at $z>5$ may also carry the imprint of cosmic re-ionization, therefore shining light on the physics of the dark ages (e.g., \citealt{Ferrara_enrich,Castellano,Ma2018}).

Crucially, while abundant optical and near infrared (NIR) observations have revealed the end product of early galaxy formation (e.g., \citealt{Bouwens2015}), the drivers of such evolution are the gas processes of intense gas inflows, outflows, and cooling which lead to primordial star formation, galaxy growth, and dynamical assembly (e.g., \citealt{Dave2011,Hopkins14}).
Such gas flows are difficult to observe directly, but measurements of the gas conditions provide the most direct constraints on the physics of early galaxy evolution. For example, the observable gas phase metallicity  probes the balance between gas inflows, outflows and metal enrichment due to star formation (e.g., \citealt{Tremonti2004,Dave2012,Lilly13}). On the other hand, the relationship between local gas properties and star formation rate (the ``star formation law") in early, forming galaxies provides the critical link between observable stellar properties and the more fundamental properties of the interstellar gas (e.g., \citealt{Sharda2018,Krumholz2018}). Since the ``star formation law" may emerge from the complex effects of stellar feedback and local gas dynamics, it is of great interest to explore its redshift evolution and any variations across galaxy types and gas conditions (e.g., \citealt{Daddi10b,Genzel15,Scoville16,Scoville17, Tacconi18,Orr2018}). A promising way to better constrain the gas metallicity and the star formation law in ``normal" galaxies is to utilize tracers of the stat-forming gas phase. In this work, we take advantage of the latest radio and (sub)millimeter interferometers to probe such tracers up to the first billion years of cosmic time.

CO rotational transitions and  atomic fine structure lines in the far infrared (FIR) provide some of the most accurate tracers of the properties of the star-forming interstellar medium (ISM) in galaxies because they are  bright, unaffected by dust extinction and probe all the main gas phases (e.g., \citealt{Stacey91,Hollenbach_review,Kaufman99,CarilliWalter}). 
In order to constrain the star formation law we need to trace the cold, molecular gas mass because it is found to be most causally connected to star formation in local galaxies \citep{Schruba11,Bigiel2011,Leroy13, CarilliWalter}. The best characterized tracers of such molecular gas are low-{\em J} rotational emission lines of the CO molecule, which have been calibrated within the Milky Way and in local galaxies, and achieve a high degree of consistency (e.g., \citealt{Leroy11, Sandstrom2013, Bolatto_review}).
These measurements may depend on metallicity estimates, since metallicity appears to strongly affect the fraction of molecular gas emitting CO lines and therefore, the gas mass-to-light ratio $\alpha_{\rm CO}$ (e.g., \citealt{MaloneyBlack,Madden97,Kaufman99, Bolatto_review}). However, it is difficult to measure metallicity directly in the cold molecular medium because no hydrogen lines are directly accessible. Indirect tracers of metallicity typically involve either probes of the nitrogen abundance ratio to other metals or probes of the hardness and intensity of the radiation field (e.g., \citealt{Masters16, Vincenzo16}). The latter technique rests on observations of local dwarf galaxies, which have shown that lower metallicity environments may produce harder and more intense ultraviolet radiation, producing stronger lines from higher ionization states (e.g., \citealt{Cormier15,Croxall17}).  Therefore, joint measurements of CO and of  metallicity probes for the same sample are of key interest to relate high redshift  observations  to the mechanisms that have been investigated and understood at low redshift.

 The [C{\sc ii}] line at 158$\mu$m is now commonly observed at high redshift as a probe of the star-forming gas and of the gas dynamics in star-forming galaxies due to its widespread distribution  (e.g., \citealt{Stacey91,Maiolino2005,Maiolino2009,Walter2009,Stacey10,RiechersHFLS3,Riechers14}).
The [C{\sc ii}]/IR luminosity ratio appears to trace the surface density of star formation, providing an important measure of starburstiness (e.g., \citealt{Luhman98,Malhotra2001}). Crucially, metallicity was shown to be the primary variable controlling the residual scatter in the [C{\sc ii}]/IR--$\Sigma_{\rm IR}$ relation \citep{Smith17}. 
However, the [C{\sc ii}] line can originate from gas where hydrogen is  ionized, neutral or molecular. Therefore observations of additional diagnostic lines that probe specific phases of the ISM are required to connect observations to physical conditions. In particular, the [N{\sc ii}] line at $205\,\mu$m is expected to be emitted under similar conditions of radiation intensity and gas density to [C{\sc ii}], but uniquely from the ionized phase (e.g., \citealt{Oberst2006,DecarliNII,Pavesi16,Tanio17}), thereby assessing the fraction of [C{\sc ii}] coming from the ionized rather than neutral gas. The [C{\sc ii}]/[N{\sc ii}] line ratio has been proposed as a metallicity tracer due to its sensitivity to abundance ratios \citep{Nagao11,Nagao12} and especially due to its sensitivity to the hardness of the radiation field  as traced by the ionization state of carbon and nitrogen in the ionized gas \citep{Cormier15,Pavesi16}. \cite{Croxall17} have now conclusively demonstrated a strong correlation between gas-phase metallicity and the [C{\sc ii}]/[N{\sc ii}] line ratio using a sample of local galaxies.

Few direct observations of the ISM in galaxies at $z>5$ are available, and the most luminous galaxies have almost exclusively been targeted to date, in particular quasar hosts and the brightest dusty star forming galaxies (DSFGs, \citealt{Maiolino2005,Maiolino2009,Walter2009,Walter12,RiechersHFLS3,Riechers14,Gullberg15,Strandet2017}). Although their brightness allows a great level of detail (e.g., \citealt{RiechersHFLS3}), it is unlikely that the conditions in the most extreme outliers are representative of typical galaxies. For example, although the fraction of dust obscured star formation in extreme starbursts is close to unity, and their metallicity may be close to solar (e.g., \citealt{Magdis2011}),  the first ALMA sample study of [C{\sc ii}] at 158$\mu$m and dust emission from normal galaxies at $z>5$ and found lower dust emission than expected  \citep{Riechers14,C15, Barisic17, Faisst17}.  We have conducted these studies of galaxies in the parent sample which is constituted of ``typical" (i.e., $\sim L^*_{UV}$) galaxies with $M_*\sim10^{10}\,M_\odot$ selected from a representative spectroscopic sample containing galaxies in various evolutionary stages selected as Lyman Break Galaxies (LBGs) or Lyman Alpha Emitters (LAEs) (i.e., the two most common selection techniques at $z>5$).  Since the ultraviolet luminosity of these galaxies is near the characteristic luminosity at $z\sim5-6$, and as they lie near the star-forming Main Sequence (e.g., \citealt{Speagle14}) as shown by \cite{Faisst16a}, we refer to these galaxies as ``normal" in the following. Their properties significantly differ from massive, hyper-luminous starbursts and quasars which have been studied in most detail in ISM studies at $z>5$ to-date, and which are typically characterized by $\sim5-10\times$  higher star formation rates (e.g., \citealt{Riechers14,Decarli2017}).
 While the ultraviolet luminosity and stellar mass of all sample galaxies is approximately equal, one of the main results of our initial ALMA observations was the wide range of [C{\sc ii}] and dust continuum luminosity observed  \citep{C15}. This wide range of FIR properties already in this small sample may suggest an evolutionary sequence, spanning the range from younger galaxies during their first major starburst to more ``mature" and dust-rich galaxies bridging the gap to, traditionally sub-mm selected, DSFGs (e.g., \citealt{C15,Pavesi16,Faisst17}). This interpretation is supported by an analysis of the IRX/$\beta_{\rm UV}$ relation \citep{Faisst17}, which found similar conditions as observed in massive galaxies at lower redshift in some galaxies while suggesting different dust properties (such as those observed in low metallicity dwarfs) for  others.

In order to constrain the conditions for star formation, low-{\em J} CO transitions provide the best probe and the most direct comparison to lower redshift surveys (e.g., \citealt{Daddi10a,Tacconi13, Tacconi18}). 
The cold molecular gas properties in ``normal" star-forming galaxies  are poorly constrained beyond $z\sim3$. Even at $z\sim3-4$ only few significant detections have been achieved, mostly afforded by strong gravitational lensing \citep{Coppin2007,Riechers10b,DZ15,DZ17}, a serendipitous detection at $z\sim3.22$ (\citealt{Avani,Avani19}), and constraining upper limits for unlensed targets (e.g., \citealt{Tan13}). The low detection rate could suggest a strong evolution in $\alpha_{\rm CO}$ with redshift, possibly driven by a rapid metallicity evolution \citep{Tan13,Tan14}.  
However, as shown by \cite{C15}, standard selection techniques at $z>5$  yield a wide range of dust obscuration, which may suggest that a corresponding range of CO enrichment may also exist.
We here present NSF's Karl G. Jansky Very Large Array (VLA) observations of the CO(2--1) transition from the FIR-brightest ``normal" galaxy of the \cite{C15} sample and of another, $\sim3$ times less FIR-bright, to obtain the first solid constraints at $z\sim6$.
 We also discuss  new ALMA measurements of the [N{\sc ii}] line luminosity in all the galaxies from the \cite{C15} sample with dust continuum and  [C{\sc ii}] detections.

In Section~2, we describe new VLA observations of the CO(2--1) line transition from HZ10 ($z\sim5.7$), and those covering LBG-1 ($z\sim5.3$; also named HZ6; \citealt{Riechers14,C15}) which were initially obtained as part of the CO Luminosity Density at High-$z$ (COLDz) survey \citep{Pavesi_COLDz}. We also present new ALMA observations targeting the [N{\sc ii}] transition at $205\,\mu$m from the dust-detected sub-sample among those presented by \cite{C15}, composed of HZ4, LBG-1, HZ9 and HZ10, expanding our previous sample study \citep{Pavesi16}.
In Section~3 we present the results from the analysis of our CO and [N{\sc ii}] measurements. In Section~4 we discuss the implications of our measurements for the metallicity, the state of maturity of the star-forming ISM and the ``star formation law"  of this sample of ``normal" galaxies at $z>5$. Finally, we present our conclusions in Section~5.  In this work, we adopt  a Chabrier IMF and a flat, $\Lambda$CDM cosmology with $H_0=70\,$km s$^{-1}$ Mpc$^{-1}$ and $\Omega_{\rm M}=0.3$.

\begin{figure*}[htb]
\centering{
 \includegraphics[width=\textwidth]{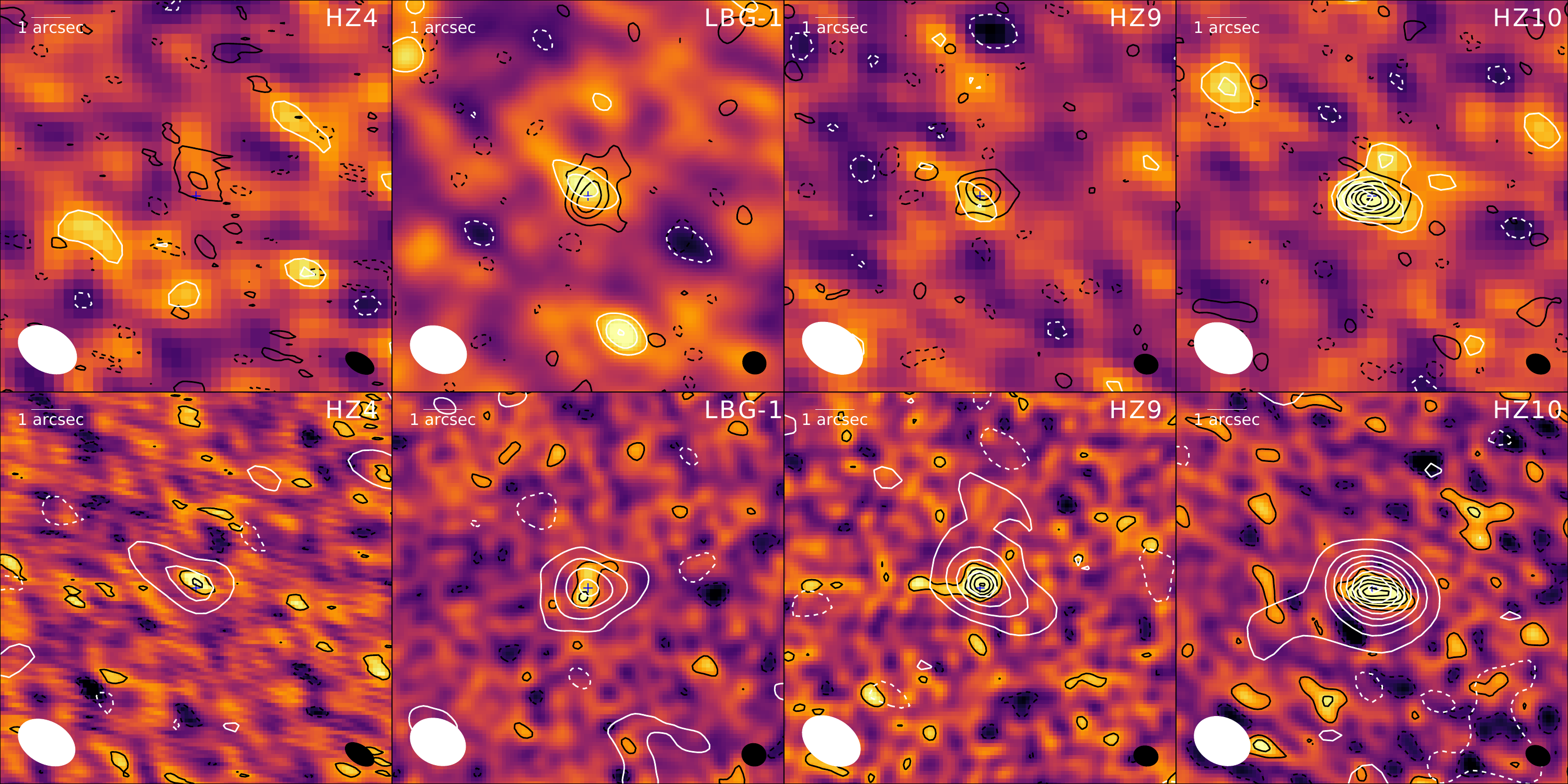}
}
\caption{Top: Integrated line maps (over the line FWHM) showing [N{\sc ii}] color-scale with  [N{\sc ii}] (white) and [C{\sc ii}] (black) contours overlaid. Blue crosses indicate the positions of the $205\,\mu$m continuum peak. The  [N{\sc ii}] ([C{\sc ii}]) beam is shown in the bottom left (right) corner of each panel. The  [N{\sc ii}] ([C{\sc ii}]) contours are multiples of $1\sigma$ ($4\sigma$), starting at $\pm2\sigma$. The noise levels in the [C{\sc ii}] line maps are 0.07, 0.04, 0.11, 0.09 Jy km s$^{-1}$ beam$^{-1}$, and in the [N{\sc ii}] line maps are 0.019, 0.016, 0.016, 0.04 Jy km s$^{-1}$ beam$^{-1}$, respectively.
Bottom: Continuum maps showing $158\,\mu$m color-scale with $205\,\mu$m (white) and $158\,\mu$m (black) contours overlaid.  Contours start at $\pm2\sigma$ and are in steps of $2\sigma$ (with the exception of the $205\,\mu$m contours in HZ9 and HZ10, in steps of $4\sigma$). The $205\,\mu$m ($158\,\mu$m) beam is shown in the bottom left (right) corner.}
\label{fig:mom_cont}
\end{figure*}

\begin{figure*}[htb]
\centering{
 \includegraphics[width=\textwidth]{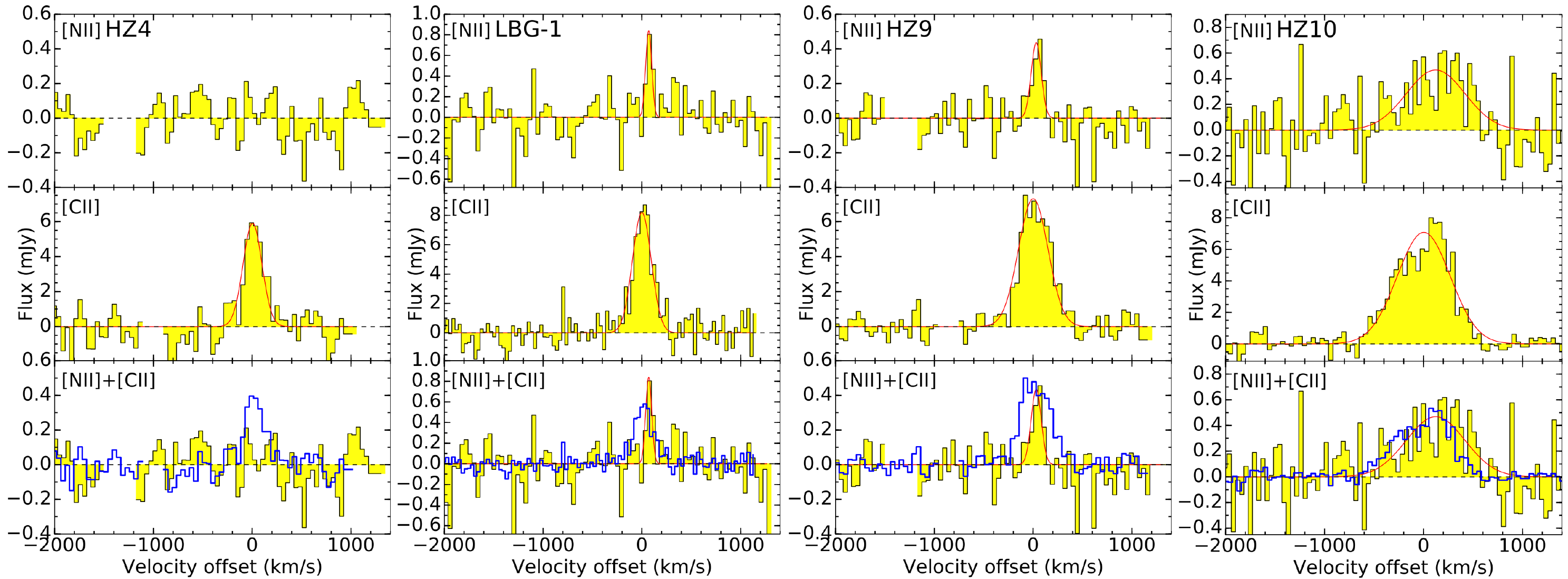}
}
\caption{[C{\sc ii}] and [N{\sc ii}] spectra of our sample galaxies, and Gaussian fits to the line emission (red curves). The channel velocity width in all spectra is $\sim42\,$km s$^{-1}$ (except in the LBG-1 [C{\sc ii}] spectra it is $\sim32\,$km s$^{-1}$). [C{\sc ii}] is scaled down by a factor 15 in flux density in the bottom panels for comparison (in blue). }
\label{fig:spectra}
\end{figure*}

\begin{figure*}[htb]
\centering{
 \includegraphics[width=.8\textwidth]{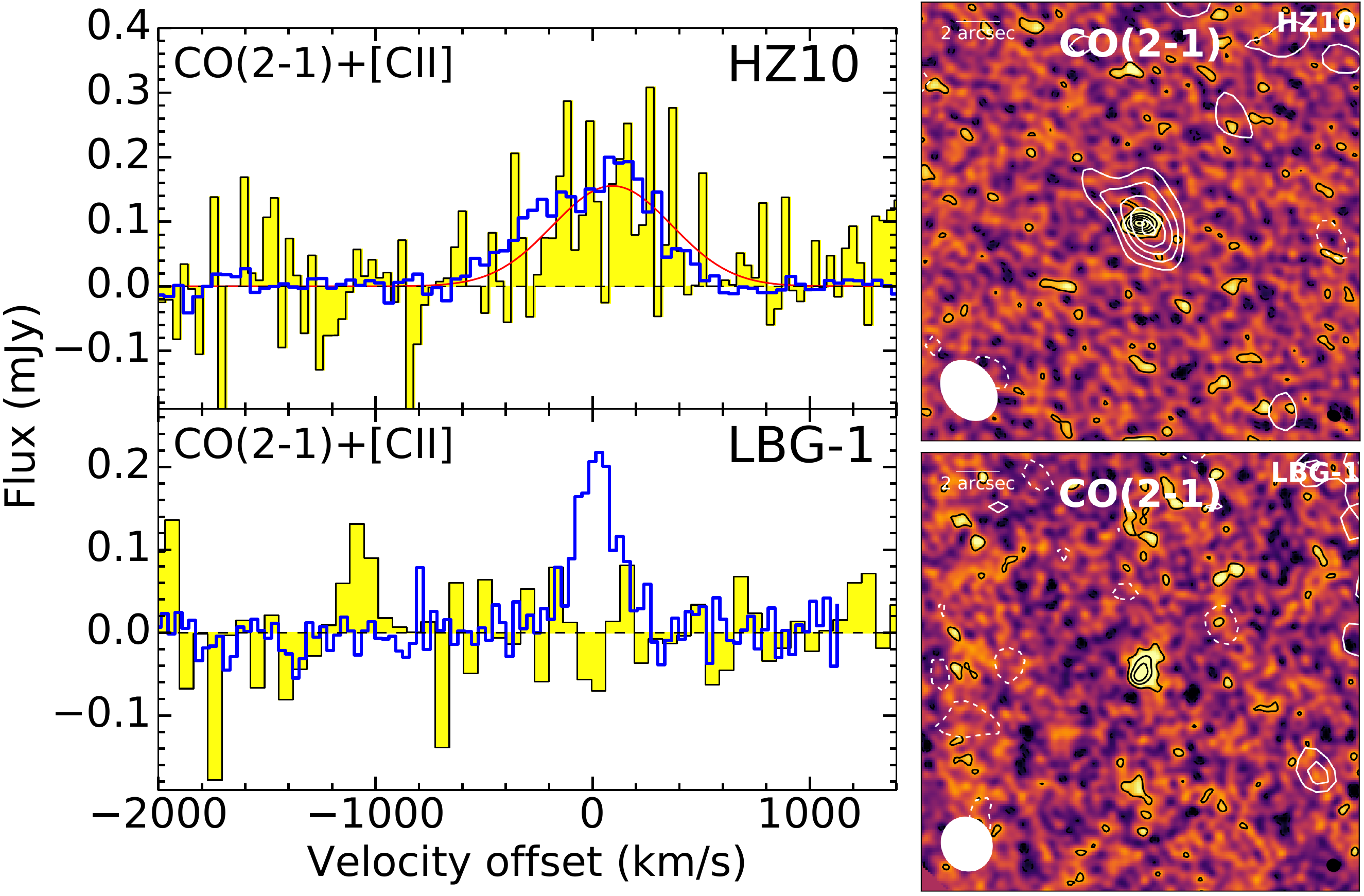} 
}
\caption{Left: CO(2--1) spectra of HZ10 and LBG-1, and Gaussian fit to the detected line emission (red curves). [C{\sc ii}] is shown scaled down by a factor 40 in flux density for comparison (in blue). The channel velocity width is $\sim40\,$km s$^{-1}$  for HZ10 in both spectra, $\sim32\,$km s$^{-1}$ and $\sim64\,$km s$^{-1}$ for LBG-1 for [C{\sc ii}] and CO, respectively. Right:  Integrated line maps (over the line FWHM) showing [C{\sc ii}] color-scale with  CO(2--1) (white) and [C{\sc ii}] (black) contours. The  CO(2--1) ([C{\sc ii}]) beam is shown in the bottom left (right) corner of each panel. The  CO(2--1) ([C{\sc ii}]) contours are multiples of $1\sigma$ ($4\sigma$), starting at $\pm2\sigma$. The noise levels in the [C{\sc ii}] line maps are 0.04, 0.09 Jy km s$^{-1}$ beam$^{-1}$ for LBG-1 and HZ10, respectively, and in the CO(2--1) line maps are 0.006, 0.010 Jy km s$^{-1}$ beam$^{-1}$ for LBG-1 and HZ10, respectively.}
\label{fig:CO_spectra}
\end{figure*}

\begin{table*}[htb]
\caption[]{Measured CO, [C{\sc ii}] and [N{\sc ii}] line properties of our sample galaxies}
\label{table_lines}
\resizebox{2\columnwidth}{!}{%
\begin{tabular}{l c c c c c c c c}
\hline 
Quantity & HZ4&LBG-1&HZ9 & HZ10 \\
\hline \noalign {\smallskip}
[C{\sc ii}] line properties\\
$\nu_{{\rm obs}}$(GHz) &$290.400\pm0.013$ &$301.980\pm0.007$&$290.545\pm0.019$&$285.612\pm0.013$\\
Redshift& $5.5445\pm0.0003$ &$5.29359\pm0.00015$&$5.5413\pm0.0004$&$5.6543\pm0.0003$ \\
$S_{{\rm [CII]}}$(mJy) &$5.9\pm0.7$&$8.2\pm0.5$&$7.3\pm0.9$&$7.1\pm0.3$\\
$FWHM_{\rm [CII]}$ (km s$^{-1}$) & $230\pm30$&$230\pm20$&$350\pm50$&$630\pm30$  \\
$I_{\rm [CII]}$ (Jy km s$^{-1}$) &  $1.3\pm0.3$&$2.1\pm0.2$&$2.7\pm0.3$&$4.5\pm0.3$\\
$L_{\rm [CII]}$ ($10^{9} L_\odot$)&$1.1\pm0.3$&$1.71\pm0.16$&$2.2\pm0.2$&$4.0\pm0.3$\\
deconvolved size & $(1^{\prime\prime}.1\pm0^{\prime\prime}.3) \times (0^{\prime\prime}.6\pm0^{\prime\prime}.3)$ &$(1^{\prime\prime}.00\pm0^{\prime\prime}.12) \times (0^{\prime\prime}.57\pm0^{\prime\prime}.10)$&$(0^{\prime\prime}.68\pm0^{\prime\prime}.12) \times (0^{\prime\prime}.48\pm0^{\prime\prime}.11)$&$(0^{\prime\prime}.80\pm0^{\prime\prime}.07) \times (0^{\prime\prime}.42\pm0^{\prime\prime}.06)$ \\ 
size (kpc$^2$) &$(6.6\pm1.8) \times (3.6\pm 1.8)$&$(6.2\pm0.7) \times (3.5\pm 0.6)$&$(4.1\pm0.7) \times (2.9\pm 0.7)$&$(4.8\pm0.4) \times (2.5\pm 0.4)$ \\
$S_{\rm 158\mu m}$ (mJy)&$0.24\pm0.05$&$0.26\pm0.07$&$0.60\pm0.09$&$1.18\pm0.16$\\
\hline
[N{\sc ii}] line properties\\
$\nu_{{\rm obs}}$(GHz) &---& $232.114\pm0.007$ &$223.348\pm0.009$&$219.49\pm0.04$\\
$S_{{\rm [NII]}}$ (mJy) & ---& $0.8\pm0.2$&$0.4\pm0.1$&$4.7\pm0.8$\\
$FWHM_{\rm [NII]}$ (km s$^{-1}$) & ---$^a$&$73\pm19$ &$120\pm30$&$700\pm130$\\
$I_{\rm [NII]}$ (Jy km s$^{-1}$) & $<0.06$&$0.06\pm0.02$&$0.05\pm0.02$&$0.34\pm0.10$\\
$L_{\rm [NII]}$ ($10^{9} L_\odot$)& $<0.04$&$0.036\pm0.012$&$0.032\pm0.013$&$0.22\pm0.07$\\
$S_{\rm 205\mu m}$ (mJy)&$0.10\pm0.02$&$0.20\pm0.03$&$0.33\pm0.04$&$0.83\pm0.05$\\
\hline
$L_{\rm [CII]}/L_{\rm [NII]}$ & $>24$& $41^{+20}_{-10}$&$61^{+40}_{-17}$&$17^{+7}_{-4}$\\
\hline
CO(2--1) line properties \\
$\nu_{{\rm obs}}$(GHz) &---&---&---&$33.157\pm0.006$\\
$S_{{\rm CO}}$ (mJy) &---&---&---&$0.16\pm0.03$\\
$FWHM_{\rm CO}$ (km s$^{-1}$) &---&---$^a$&---&$650\pm140$\\
$I_{\rm CO}$ (Jy km s$^{-1}$) &---&$<0.018$&---&$0.10\pm0.02$\\
$L'_{\rm CO}$ ($10^{10}$ K km s$^{-1}$ pc$^2$)&---&$<0.44$&---&$2.9\pm0.6$\\
$S_{\rm 34 GHz}$ ($\mu$Jy)&---&$<4$&---&$<7.8$\\
\hline \noalign {\smallskip}
\end{tabular}
}
\tablecomments{All  quoted uncertainties correspond to $1\sigma$ statistical uncertainty intervals and all limits correspond to $3\sigma$. $^a$: We assume FWHM equal to that of the [C{\sc ii}] line in order to derive upper limits on the line flux.}
\end{table*}

\section{Observations}
\subsection{VLA observations of CO(2--1)}
We observed the CO(2--1) transition in HZ10 using the VLA in Ka band (project ID: 17A-011, PI: Pavesi). A complete description of these observations may be found in \cite{CRLE}, which describes the properties of CRLE, an hyper-luminous DSFG at the same redshift as HZ10 and located within the same field view, with a separation of only 13$^{\prime\prime}$. In three of the eight observing sessions the two intermediate frequencies (IFs) were tuned to the central frequency of the CO(2--1) line in HZ10, and to the adjacent frequency range to maximize continuum sensitivity. In the remaining five sessions the second IF was moved  in order to provide uninterrupted coverage of the CO(2--1) line in CRLE, by partially overlapping the first IF \citep{CRLE}. The total observing time was 19.8 hrs, on source.
We imaged the data with the {\sc clean} algorithm in the Common Astronomy Software Application ({\sc casa}  version 4.7, using natural weighting for maximal sensitivity. The imaging of the CO line data results in a synthesized beam size of $3.0^{\prime\prime}\times2.3^{\prime\prime}$ at the redshifted CO(2--1) frequency and $2.7^{\prime\prime}\times2.3^{\prime\prime}$ in the continuum map. The rms noise at the position of HZ10 (i.e., at the phase center) is $\sim45\,\mu$Jy beam$^{-1}$ in a 35 km s$^{-1}$ wide channel. The final rms noise when averaging over the line-free 2.0 GHz of bandwidth is $\sim2.7\,\mu$Jy beam$^{-1}$. 
The CO(2--1) transition in LBG-1 was observed as part of the COLDz survey \citep{Pavesi_COLDz,Riechers_COLDz}, and a preliminary version was shown by \cite{Riechers14}. A complete description of these observations and of the imaging may be found in \cite{Pavesi_COLDz}. The equivalent time on-source in the mosaic is 14 hours at the position of LBG-1. The imaging of the CO line data results in a synthesized beam size of $2.5^{\prime\prime}\times2.3^{\prime\prime}$ at the redshifted CO(2--1) frequency and $2.7^{\prime\prime}\times2.4^{\prime\prime}$ in the continuum map. The rms noise at the position of LBG-1 is $\sim67\,\mu$Jy beam$^{-1}$ in a 35 km s$^{-1}$ wide channel. The final rms noise when averaging over the full 8 GHz of bandwidth is $\sim1.3\,\mu$Jy beam$^{-1}$.

\subsection{ALMA observations of [C{\sc ii}] and [N{\sc ii}]}
%CII observing details for HZ4 and 9 (the stuff I reported in 2016)
%for the other two say it's same reduction at 2016
Our observations of the [C{\sc ii}] line, data reduction and imaging for LBG-1 and HZ10 were previously described by \cite{Riechers14,C15}, and \cite{Pavesi16}.
The ALMA Cycle-1 observations targeting the [C{\sc ii}] lines for HZ4 and HZ9, have previously been presented  by \cite{C15}  and here we provide a brief description of the data that we have re-processed and re-analyzed.
These observations were taken on 2013 November 4-16 in band 7 as part of a larger project (ID: 2012.1.00523.S, PI: Capak). The HZ4 pointing resulted in $20\,$min on source with 28 usable antennae. Ganymede was observed as flux calibrator, J0522$-$3627 was observed as bandpass calibrator, and J1008$+$0621 was observed as amplitude/phase gain calibrator.
The  HZ9 data resulted in $38\,$min on source with 27 antennae. Ganymede was observed as flux calibrator, J1037$-$2934 was observed as bandpass calibrator, and J1058$+$0133 was observed as amplitude/phase gain calibrator.
In both cases the correlator was set up to target the expected frequency of the [C{\sc ii}] line and to provide continuous coverage of the continuum emission in adjacent spectral windows with channels of $15.6\,$MHz in Time Division Mode (TDM).
 {\sc casa} version 4.5 was used for data reduction and analysis. All images and mosaics were produced with the {\sc clean} algorithm, using natural weighting for maximal sensitivity.
For HZ4, the imaging results in a synthesized beam size of $0.8^{\prime\prime} \times 0.5^{\prime\prime}$ at the redshifted  [C{\sc ii}] frequency and in the continuum map. The rms noise in the phase center is $\sim0.5\,$mJy beam$^{-1}$ in a $44\,$km s$^{-1}$ wide channel and the final rms noise when averaging over all spectral windows (i.e. over a total $7.5\,$GHz of bandwidth) is $\sim 54\, \mu$Jy beam$^{-1}$.
For HZ9, the imaging results in a synthesized beam size of $0.6^{\prime\prime} \times 0.5^{\prime\prime}$ at the redshifted  [C{\sc ii}] frequency and in the continuum map. The rms noise in the phase center is $\sim0.4\,$mJy beam$^{-1}$ in a $43\,$km s$^{-1}$ wide channel and the final rms noise when averaging over all spectral windows (i.e. over a total $7.5\,$GHz of bandwidth) is $\sim 47\, \mu$Jy beam$^{-1}$.

Cycle-3 observations of [N{\sc ii}] 205$\,\mu$m targeting our sample galaxies were taken on 2016 January 1 and 5 in band 6, as part of two separate programs (2015.1.00928.S and 2015.1.00388.S, PIs: Pavesi and Lu, respectively) with one track from each program for HZ10 and LBG-1 and one track for HZ4 and HZ9 from the second program, taken in a compact configuration (max. baseline $\sim300\,$m). Observations from the first program were previously described by \cite{Pavesi16}, and the HZ10 observations for both programs were previously described by \cite{CRLE}. We here present the remaining observations for LBG-1, HZ4 and HZ9.
The two sets of observations for LBG-1 resulted in 64 min and  18 min on source respectively, with $\sim$41--45 usable $12\,$m antennae under good weather conditions at 1.3 mm.
The first set of observations was previously described by \cite{Pavesi16}. 
For the second set of observations of LBG-1, the nearby radio quasar J0948$+$0022 was observed regularly for amplitude and phase gain calibration, and J0854$+$2006 was observed for bandpass and flux calibration. 
The observations of HZ4 and HZ9 resulted in 30 and 47 min on source, with 45 and 47 usable $12\,$m antennae, respectively. The same radio quasar was observed for amplitude and phase calibration as for LBG-1, and J1058+0133 was observed for bandpass and flux calibration.
The correlator was set up in an identical configuration for these observations, to cover two spectral windows of 1.875 GHz bandwidth each at $15.6\,$MHz ($\sim20\,$km s$^{-1}$) resolution (dual polarization) in Time Division Mode (TDM), in each sideband. We estimate the overall accuracy of the flux calibration to be within $\sim10\%$.
We used {\sc casa} version 4.5 for data reduction and analysis.  We combined data from all observations, and produced all images with the {\sc clean} algorithm, using natural weighting for maximal point source sensitivity.
 Imaging the [N{\sc ii}] data for HZ4 results in a synthesized beam size of $1.6^{\prime\prime} \times 1.1^{\prime\prime}$ at the redshifted  [N{\sc ii}] frequency of HZ4 and in the continuum map. The rms noise in the phase center is $\sim0.14\,$mJy beam$^{-1}$ in a $44\,$km s$^{-1}$ wide channel.  The final rms noise when averaging over the line free spectral windows (i.e. over a total $7.5\,$GHz of bandwidth) is $\sim 13\, \mu$Jy beam$^{-1}$. 
  Imaging the [N{\sc ii}] data for LBG-1 results in a synthesized beam size of $1.5^{\prime\prime} \times 1.2^{\prime\prime}$ at the redshifted  [N{\sc ii}] frequency of LBG-1 and in the continuum map. The rms noise in the phase center is $\sim0.16\,$mJy beam$^{-1}$ in a $40\,$km s$^{-1}$ wide channel.  The final rms noise when averaging over the line free spectral windows (i.e. over a total $7.5\,$GHz of bandwidth) is $\sim 15\, \mu$Jy beam$^{-1}$.   Imaging the [N{\sc ii}] data for HZ9 results in a synthesized beam size of $1.7^{\prime\prime} \times 1.2^{\prime\prime}$ at the redshifted  [N{\sc ii}] frequency of HZ9 and in the continuum map. The rms noise in the phase center is $\sim0.15\,$mJy beam$^{-1}$ in a $44\,$km s$^{-1}$ wide channel.  The final rms noise when averaging over the line free spectral windows (i.e. over a total $7.5\,$GHz of bandwidth) is $\sim 14\, \mu$Jy beam$^{-1}$. 
   Imaging the [N{\sc ii}] data for HZ10 results in a synthesized beam size of $1.6^{\prime\prime} \times 1.2^{\prime\prime}$ at the redshifted  [N{\sc ii}] frequency of HZ10 and in the continuum map. The rms noise in the phase center is $\sim0.14\,$mJy beam$^{-1}$ in a $44\,$km s$^{-1}$ wide channel.  The final rms noise when averaging over the line free spectral windows (i.e. over a total bandwidth of $7.5\,$GHz) is $\sim 19\, \mu$Jy beam$^{-1}$ (Figures~\ref{fig:mom_cont} \& \ref{fig:spectra}).

\begin{table}[htb]
\caption[]{Derived properties of our sample galaxies}
\label{table_derived}
\resizebox{\columnwidth}{!}{%
\begin{tabular}{l c c c c c c c c}
\hline 
Quantity & HZ4&LBG-1&HZ9 & HZ10 \\
\hline \noalign {\smallskip}
$L_{\rm FIR}$ ($10^{11}\,L_\odot$)&$5.2^{+4.6}_{-2.6}$ &$4.9^{+4.4}_{-2.6}$&$12^{+10}_{-6}$&$13^{+11}_{-7}$\\
SFR ($M_\odot$ yr$^{-1}$) &$52^{+46}_{-26}$ &$49^{+44}_{-26}$&$120^{+100}_{-60}$&$130^{+110}_{-70}$\\
$M_*$ ($10^{9}\,M_\odot$)$^a$&$4.7^{+2.9}_{-1.8} $&$15^{+6}_{-5}$ &$7.2^{+5.0}_{-2.9}$&$25^{+12}_{-8} $\\
$M_{\rm gas}$ ($10^{10}\,M_\odot$)$^b$&---&$<2$&---&$13\pm3$\\
$M_{\rm dyn}(<R_{1/2})$ ($10^{10}\,M_\odot$)&$1.8^{+1.3}_{-1.0}$&$1.9^{+0.6}_{-0.4}$&$3.5^{+3.1}_{-1.6}$&$10\pm3$\\
\hline \noalign {\smallskip}
\end{tabular}
}
\tablecomments{All  quoted uncertainties correspond to $1\sigma$ intervals and all limits correspond to $3\sigma$.
$^a$: Stellar masses reported by \cite{C15}.
$^b$: Gas masses are derived from the CO luminosity assuming a Galactic $\alpha_{\rm CO}\sim4.5$ conversion factor.}
\end{table}

\section{Analysis}

\subsection{Dust continuum measurements}

We detect dust continuum emission from the full galaxy sample at 158$\,\mu$m and 205$\,\mu$m (Figure~\ref{fig:mom_cont}, Table~\ref{table_lines}). 
No continuum signal is detected in the VLA observations targeting HZ10 and LBG-1 at $\sim$34 GHz (corresponding to rest-frame $\sim1.3\,$mm), yielding deep $3\sigma$ upper limits (Table~\ref{table_lines}).
We measure the continuum flux at 158$\,\mu$m and 205$\,\mu$m by imaging all line-free channels using natural baseline weighting, and using the {\sc casa} task {\sc imfit} to fit a 2D Gaussian model to the emission. 
 These detections and upper limits represent the only available constraints to the FIR SED and we use them in the following to constrain the FIR luminosity and to provide initial gas mass estimates through the Rayleigh-Jeans method.

 We follow standard procedure and fit these continuum fluxes with a modified black-body smoothly connected to a mid-IR power law (e.g., \citealt{Casey12,Riechers14,Pavesi16,Faisst17}).   The results of the FIR SED fitting, together with the optical-to-FIR SED for the full galaxy sample, are shown in Appendix~A. We adopt a Bayesian approach and employ a Markov Chain Monte-Carlo technique through \texttt{emcee} to infer the posterior distribution for the modified black-body parameters: dust temperature, dust emissivity $\beta$ parameter, mid-IR power law index, flux normalization and wavelength at which the optical depth equals unity. We adopt high dust temperature priors, as suggested by \cite{Faisst17} for these galaxies. We employ Gaussian priors for the dust emissivity $\beta$ parameter ($1.7\pm0.5$), for the dust temperature ($60\pm15$ K), for the mid-IR power law index ($2.0\pm0.5$) and for the transition rest-frame wavelength to the optically thick regime ($60\pm20\,\mu$m). We note that the relative fluxes at 158$\,\mu$m and 205$\,\mu$m across our sample  (with ratios ranging from $1.3\pm0.4$ to $2.4\pm0.8$) suggest a diversity of dust SED shapes. 
We derive far-infrared (FIR) luminosities by integrating between 42.5 and 122.5 $\mu$m (Table~\ref{table_derived}). 
The available constraints are not sufficient to  completely resolve the degeneracy between dust temperature and emissivity index variations,  which are however fully captured by our Bayesian approach and contribute to the uncertainties quoted for the FIR luminosity.
Because the dust spectral energy distributions (SEDs) are not constrained in the mid-IR, we follow the standard practice of adopting the FIR luminosities as an estimate of total IR, without extrapolating to shorter IR wavelengths (e.g., \citealt{Riechers14,CRLE}). We caution, however, that this may be an under-estimate, and that the total IR luminosity may be $\sim$1.5-2$\times$ higher than the FIR alone. 

%We can use the available dynamical mass and Rayleigh-Jeans dust continuum emission estimates to provide constraints to the gas masses in these galaxies, independently from the CO measurements.
In order to provide constraints on the gas masses in these galaxies independently from the CO measurements, we can use the Rayleigh-Jeans dust continuum emission. This  will  provide the first constraints to the $\alpha_{\rm CO}$ conversion factor in ``normal" galaxies at $z>3$ in the following.
The Rayleigh-Jeans dust continuum emission has been used to estimate dust and gas masses, assuming an average emissivity and dust temperature for the dominant cold dust component, and a constant dust-to-gas ratio (\citealt{Hildebrand,Eales12,Scoville_book,Scoville13,Scoville16,Scoville17,Bourne13,Groves15}).
The dependence on cold dust temperature and dust-to-gas ratio may make the Rayleigh-Jeans method less reliable than at lower redshifts (e.g., \citealt{CRLE}). On the other hand, the opposing effects of increasing dust temperatures, and decreasing dust-to-gas ratios that may occur in ``normal" galaxies at high redshift may partially compensate each other, as also found in  recent simulations that are consistent with this approach to gas mass measurement (e.g., \citealt{Privon18,Liang18}).
We here adopt Equations~10 and 13 in \cite{Scoville16} to derive gas mass estimates based on our continuum flux measurements, through the same assumptions which were used in those lower redshift samples \citep{Scoville16,Scoville17}.
The 34 GHz upper limits imply $3\sigma$ gas mass limits of $<2.8\times10^{11}\,M_\odot$ for HZ10 and $<1.6\times10^{11}\,M_\odot$ for LBG-1, adopting the relation derived by \cite{Scoville16,Scoville17}.
We also use the $\sim$230 GHz continuum fluxes to derive approximate estimates, although these measurements may not lie on the Rayleigh-Jeans tail, and therefore may not accurately trace the cold dust component. These continuum measurements would imply gas masses of 
$\sim1.3\times10^{10}\,M_\odot$ for HZ4, $\sim2.5\times10^{10}\,M_\odot$ for LBG-1, 
$\sim4.4\times10^{10}\,M_\odot$ for HZ9, and $\sim1.1\times10^{11}\,M_\odot$ for HZ10, with dominant systematic uncertainties due to the extrapolation of the method to very high redshift.

\begin{figure}[htb]
\centering{
  \includegraphics[width=.5\textwidth]{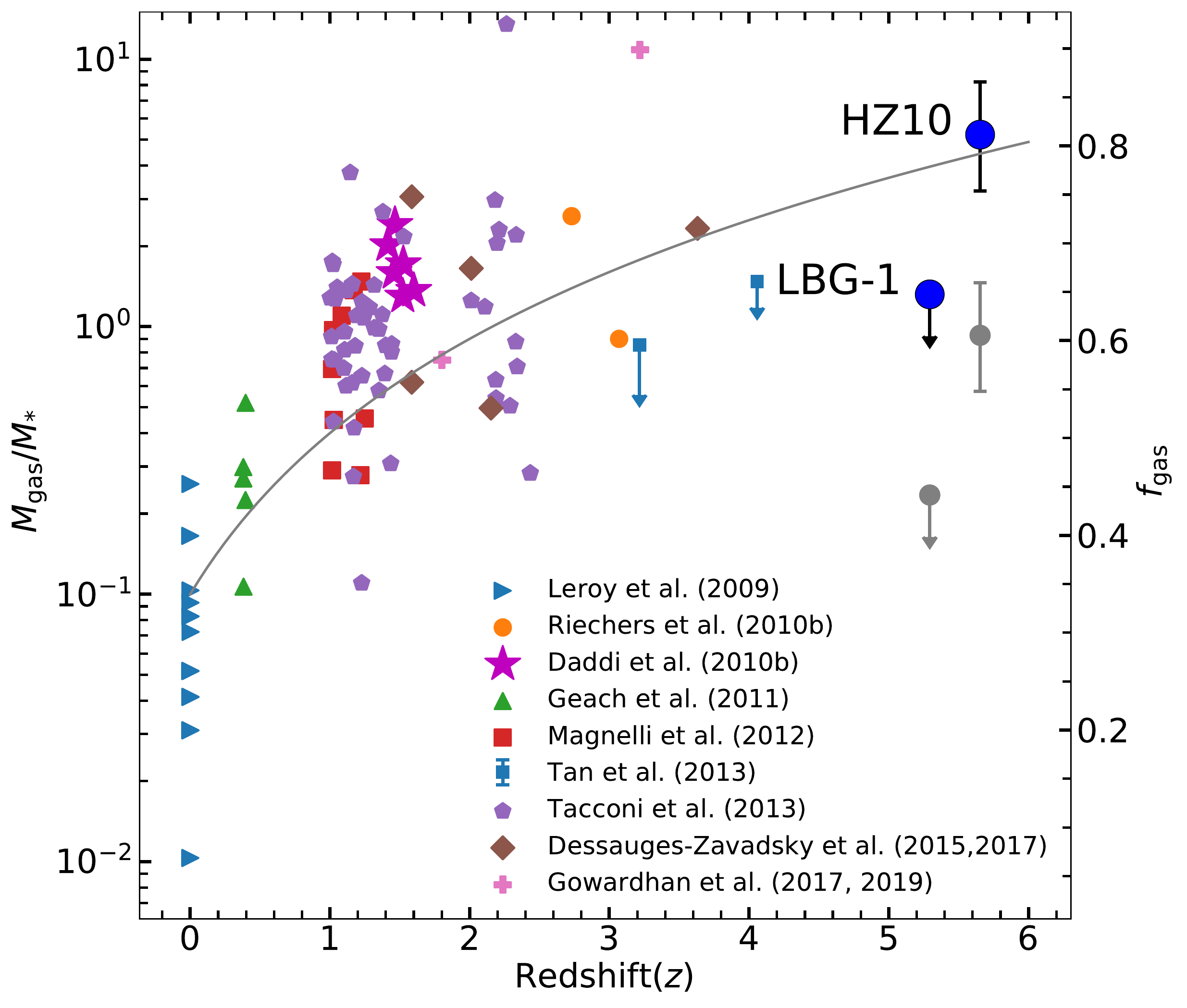}  
}
\caption{The ratio of molecular gas mass to stellar mass (calculated using an $\alpha_{\rm CO}$=4.5 $M_\odot\,({\rm K\, km\, s^{-1} \,pc^2})^{-1}$ for all sources) adapted from \cite{CarilliWalter,Leroy09,Riechers10b,Daddi10a,
Geach2011,Magnelli2012,Tan13,Tacconi13,DZ15,DZ17,Avani,Avani19}
The grey line shows $M_{\rm gas}$/$M_*\propto (1+z)^2$ \citep{Geach2011}.
An alternative choice of $\alpha_{\rm CO}$=0.8 for HZ10 and LBG-1 is also shown in light grey. 
 Stellar masses are adopted from \cite{C15}.
}
\label{fig:gas_frac}
\end{figure}

\subsection{ CO measurements}
We detect CO(2--1) line emission from HZ10 with a significance of $\gtrsim8\sigma$, and provide a constraining upper limit to the CO(2--1) emission toward LBG-1 (Figure~\ref{fig:CO_spectra}). We extract an aperture spectrum for HZ10\footnote{\label{aper_foot}We adopt elliptical apertures of sizes equal to the FWHM of the best fit 2D Gaussian to the integrated line emission.} and a single pixel spectrum at the peak position of the [C{\sc ii}] emission toward LBG-1, in order to measure or constrain the CO(2--1) line properties (Table~\ref{table_lines}). 
The CO(2--1) emission toward HZ10 appears slightly resolved, although the coarse resolution of compact array configuration observations does not allow a precise size determination. We use {\sc casa} {\sc uvmodelfit} to fit a circular Gaussian model to the line visibilities in HZ10 and derive a deconvolved FWHM size of $1^{\prime\prime}.2\pm0^{\prime\prime}.4$ for the CO(2--1) emission, corresponding to $7\pm2$ kpc, which is compatible with the [C{\sc ii}] and [N{\sc ii}] size estimates.  We do not make use of this  size estimate in the following, but use it to validate that CO and [C{\sc ii}] emission sizes for our target are consistent within the relative uncertainties. Since we find compatible values we will adopt [C{\sc ii}] emission sizes for gas reservoir sizes in the following\footnote{This is a common assumption, since both CO and [C{\sc ii}] trace the extent of the gas distribution (e.g., \citealt{deBreuck14,Litke19})}.

Our upper limit indicates that CO(2--1) line emission from LBG-1 is unexpectedly weak, relative to low redshift trends. This may be expected as a consequence of low metallicity and relatively low dust abundance \citep{Bolatto_review}, and as previously observed at $z>1$ on multiple occasions (e.g., \citealt{Genzel12,Tan14}). LBG-1 shows an unusually high inferred [C{\sc ii}]/CO(1--0) luminosity ratio  ($\gtrsim9000$, Figure~\ref{fig:CO_spectra}), relative to the value in HZ10 ($\sim3000$) and to the values commonly measured in local starbursts and high redshift galaxies ($\sim4400$; e.g., \citealt{Wolfire89,Stacey91,Stacey10}) and to a sample of high redshift dusty star forming galaxies ($5200\pm1800$; e.g., \citealt{Gullberg15}). The high ratios observed in LBG-1  cannot be explained within standard PDR models, but they naturally arise as a consequence of lower metallicity (e.g., \citealt{MaloneyBlack,Stacey91,Madden97}). In particular, low metallicity dwarf galaxies typically show similar ratios of $\sim7000-10^5$ (e.g., \citealt{Cormier14,Jameson2018}).
On the other hand, the  normal ratio observed in HZ10 points to star-forming gas properties that are similar to what is observed in lower redshift Main Sequence galaxies.

The CO(2--1) line luminosity is expected to provide a reliable estimate of the molecular gas mass. However, several factors affect the proportionality factor such as heating from,  and contrast against, the cosmic microwave background (CMB; e.g., \citealt{daCunhaCMB}) and  the strong metallicity dependence of the CO luminosity per unit molecular gas mass  (e.g., \citealt{Bolatto_review}).  Current samples including both dusty star-forming galaxies, and main-sequence galaxies at high redshift show nearly thermalized gas excitation up to the {\em J}=2--1 transition ($R_{21}\sim0.80-0.95$; e.g., \citealt{CarilliWalter, Daddi15}). Here, we therefore assume  a brightness temperature ratio of $R_{21}=1$ between the CO {\em J}=2--1 and 1--0 transitions. 
 CMB contributions at $z>5$ are only weakly constrained without additional CO excitation measurements. \cite{daCunhaCMB} suggest that we may expect the observed CO line flux to be suppressed by a factor $\sim1.25-2$ at this redshift. We do not attempt to estimate this effect independently, but simply absorb it into the definition of $\alpha_{\rm CO}$.

We can use the gas mass estimates based on the Rayleigh-Jeans dust continuum emission to constrain the $\alpha_{\rm CO}$ conversion factor by assuming that the gas mass is dominated by molecular gas. The main uncertainties inherent in the Rayleigh-Jeans dust method are a dependence on the gas-to-dust ratio and on dust properties affecting the dust SED.  The effects of these uncertainties dominate over the CO and dust continuum measurement uncertainties.
The  gas mass estimates presented in Section~3.1 imply constraints to $\alpha_{\rm CO}$\footnote{Units of $M_\odot\,({\rm K\, km\, s^{-1} \,pc^2})^{-1}$ assumed throughout the following} of $\lesssim10$  ($\sim4$ based on the 220 GHz flux)  for HZ10 and $\gtrsim5.7$\footnote{\label{Stats_note}In order to deal with relative uncertainties of order unity throughout this work, we adopt the convention of quoting Gaussian-equivalent percentiles. Therefore, uncertainty ranges correspond to 16th, 50th and 84th percentiles and $3\sigma$ limits are defined to imply a 99.7\% probability. Propagation of these uncertainties to derived quantities was carried out by numerical sampling and evaluation of posterior distribution percentiles. Lognormal distributions were used to sample skewed distributions described by asymmetric $1\sigma$ ranges. Upper limits from non-detections are treated as positive-truncated (enforcing a uniform prior), 0-centered Gaussians with specified standard deviation as determined by the noise level.} based on the 230 GHz flux for LBG-1.  These estimates are dominated by the systematic uncertainty inherent in extrapolating the Rayleigh-Jeans method to very high redshift where it has not yet been validated.  These approximate estimates are in agreement with our inference of ``normal" star-forming gas properties for HZ10, with $\alpha_{\rm CO}$ near the Milky Way value, and of lower metallicity gas in LBG-1, as signaled by an elevated $\alpha_{\rm CO}$.

\subsection{Dynamical mass and gas mass constraints}

In order to estimate dynamical masses for the full galaxy sample, we adopt a commonly used empirical procedure based on the line width and the gas emission size inferred from the integrated line emission which was calibrated on disk galaxy simulations (e.g., \citealt{Daddi10a}). The advantage of such method is the applicability to our full galaxy sample, and a more straightforward comparison to most dynamical mass estimates available in the literature, which are typically relying on such estimates (e.g., \citealt{Tacconi08,Forster09,Engel10,Walter12,CarilliWalter,Riechers14,C15,Venemans16,ASPECS3,Oteo2016}).

We apply this technique by using the line FWHM, the fitted half-light radii of the [C{\sc ii}] emission and the disk inclination from the ratio of minor to major axes (Table~\ref{table_derived}). 
The inferred dynamical mass for LBG-1 is $\sim2.5$ times lower than previous estimates, although within the original uncertainties, due to a revised [C{\sc ii}] size and due to differences in the method employed \citep{Riechers14}. However, dynamical mass estimates for LBG-1 may be affected by complex velocity structure due to interactions.

%{\bf We validate these dynamical mass estimates through a {\sc uv}-space dynamical modeling method detailed in Appendix~A. The [C{\sc ii}] line first moment maps of HZ9 and HZ10 show clear evidence for dynamical information (Appendix~A). Dynamical modeling of the same observations of HZ9 and HZ10 was previously attempted by \cite{Jones17} through the tilted-rings approach to constrain a rotating disk model.}
In Appendix~B we present an alternative dynamical analysis which directly models the observed visibilities\footnote{The main advantage of such visibility-based rather than image-based approaches is independence from deconvolution and imaging and from the statistical dependence of image pixels introduced by the non-local synthesized beam.}, following the method previously described by \cite{CRLE}. The inferred dynamical mass estimates are in agreement with those derived by \cite{Jones17} based on tilted-rings modeling in the image plane, and with those based on the \cite{Daddi10a} method for HZ9 and HZ10.

We can therefore use our dynamical masses to provide approximate estimates of the total gas masses in the full sample, by accounting for the contribution of stellar\footnote{ Stellar masses for all our galaxies are constrained by the rich multi-wavelength coverage available in the COSMOS field and were measured by \cite{C15} and \cite{Laigle16} for the entire parent sample and are typically uncertain to within a factor $\sim2$ (Table~\ref{table_derived}). 
The stellar mass fits include  deep (25.5 AB mag) COSMOS/SPLASH \citep{Steinhardt14} {\em Spitzer}/IRAC photometry  to probe wavelengths redder than rest-frame 4000 \AA, which is crucial for constraining stellar masses to this accuracy (e.g, \citealt{Faisst16b,Laigle16}).
However, we caution that stellar masses at $z>4-5$ are difficult to constrain because the rest-frame $1-2\,\mu$m wavelength emission, which is the most accurate tracer, will not be observable in such faint galaxies at $z>5$ until the {\em James Webb Space Telescope} becomes operational.} and  dark matter masses (25\%) following  \cite{Daddi10a}.  Using the CO line luminosity measurements and limits in HZ10 and  LBG-1, these independent gas mass estimates allow us to place the first constraints to the $\alpha_{\rm CO}$ conversion factor in normal galaxies at this redshift.
The dynamical mass estimates would imply total gas masses of $(1.4\pm0.9)\times10^{10}\,M_\odot$ for LBG-1, $4.5^{+4.5}_{-2.5}\times10^{10}\,M_\odot$ for HZ9 and $(1.2\pm0.5)\times 10^{11}\,M_\odot$ for HZ10. If we assume this gas mass to be dominated by molecular gas,  and thus, divide by the CO line luminosity constraints\footnote{Following the procedure explained in footnote~\ref{Stats_note}}, these estimates imply an $\alpha_{\rm CO}$ (in units of $M_\odot\,({\rm K\, km\, s^{-1} \,pc^2})^{-1}$) of $4.2^{+2}_{-1.7}$ for HZ10, but do not provide a significant constraint for LBG-1\footnote{Due to the large uncertainty in the gas mass estimate, the 3$\sigma$ CO limit only implies a 3$\sigma$ limit of $\alpha_{\rm CO}>0.2$ when appropriately propagated through posterior sampling. See footnote~\ref{Stats_note}.}.
The estimated $\alpha_{\rm CO}$ factor for HZ10 is compatible with the Milky Way value ($\sim4.5$, in the same units; \citealt{Bolatto_review}) which may also apply to $z\sim1-2$ main-sequence disk galaxies \citep{Daddi10a,CarilliWalter,Tacconi13,Genzel15}.  In the following, we assume a fixed value of $\alpha_{\rm CO}=4.5$ for definiteness, in order to derive and constrain gas masses in HZ10 and LBG-1 (Table~\ref{table_derived}, Figure~\ref{fig:gas_frac}), with the caveat that this value may only be a lower limit in the case of LBG-1, due to metallicity effects (e.g., \citealt{Bolatto_review}).

\subsection{Constraints to high redshift star formation}
\begin{figure*}[htb]
\centering{
  \includegraphics[width=.48\textwidth]{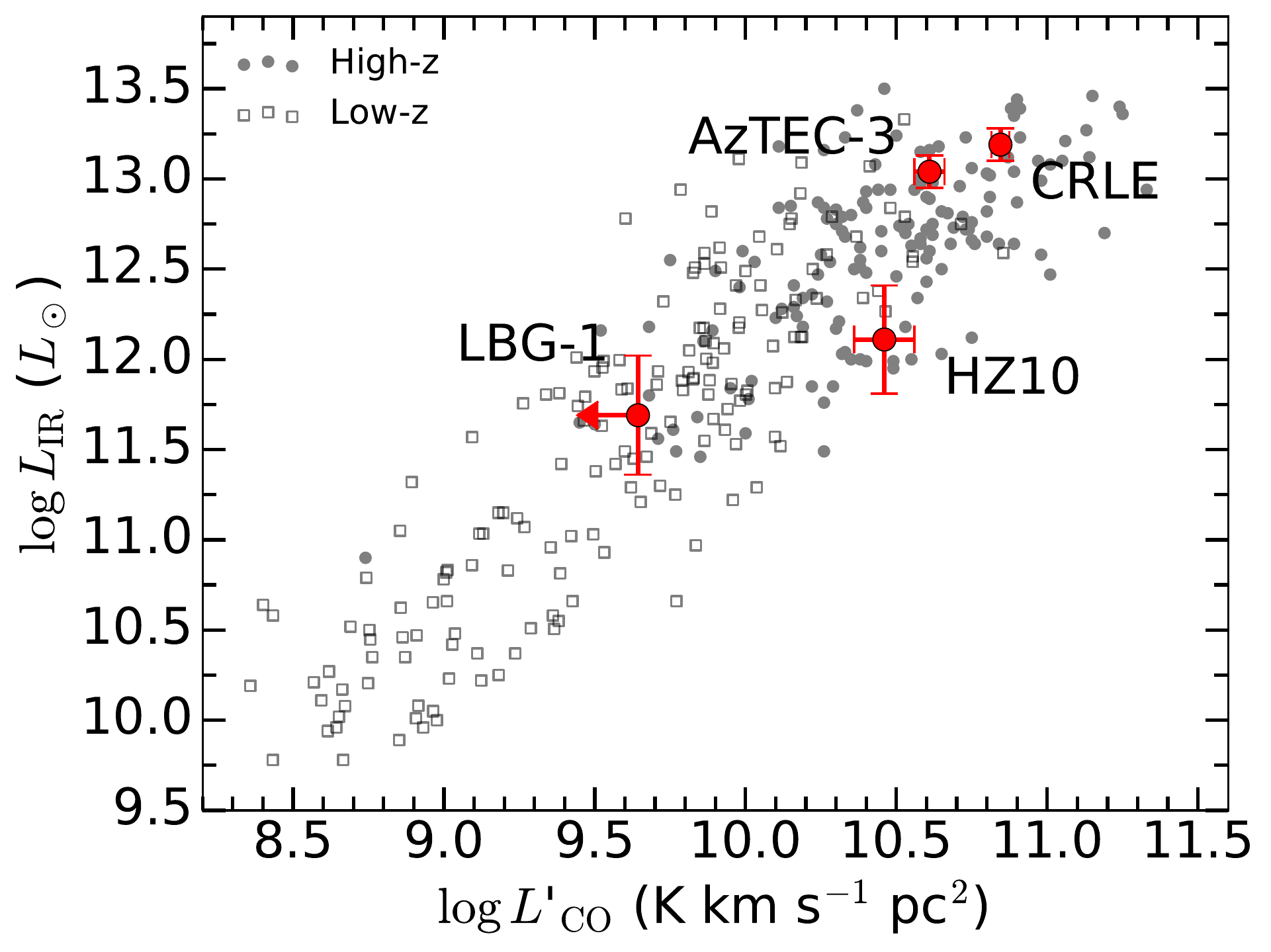}  \includegraphics[width=.48\textwidth]{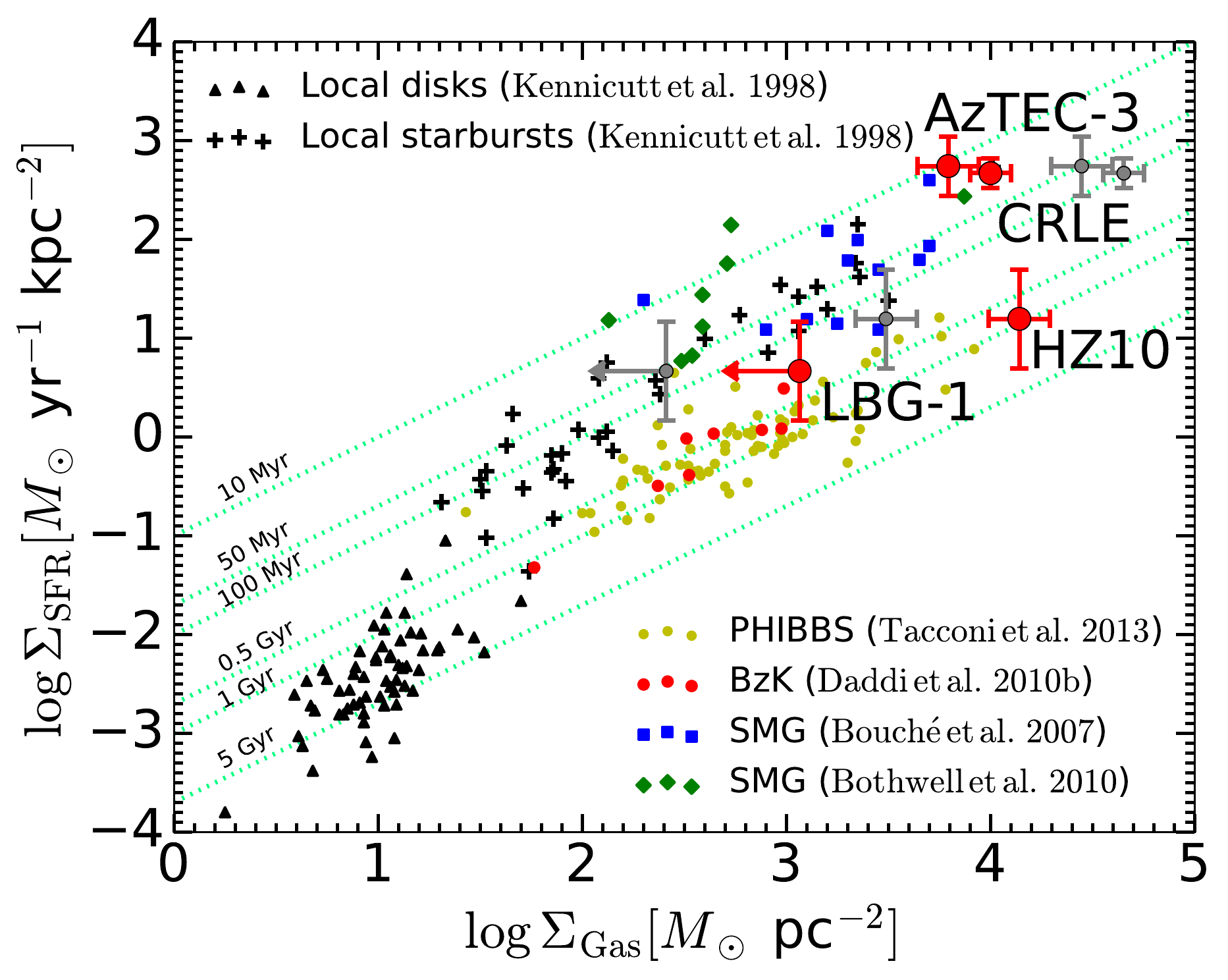} 
}
\caption{Left: IR luminosity observed in a sample of local and high redshift galaxies as a function of their CO luminosity, for comparison to the measurements in HZ10 and LBG-1 \citep{CarilliWalter}. We also include two $z>5$ DSFGs for reference (AzTEC-3 and CRLE; \citealt{Riechers10a,Riechers14,CRLE,Pavesi_COLDz}), clearly occupying a distinct part of the parameter space. 
Right: Star formation rate surface density as a function of the gas mass surface density for a sample of local and high redshift galaxies, including HZ10, LBG-1, AzTEC-3 and CRLE (adapted from \citealt{Daddi10b}, updated by \citealt{Tacconi13}). The star formation rate surface density was estimated for the $z>5$ galaxies through the FIR luminosity and the dust continuum sizes. The gas surface density was estimated uniformly for these galaxies through the CO luminosity and the [C{\sc ii}] emission spatial size,  which is compatible with our current constraints on the CO emission size. Following \cite{Daddi10b} we adopt $\alpha_{\rm CO}\sim4.5$ for Main Sequence galaxies such as HZ10 and LBG-1 and the other disk galaxies,  and $\alpha_{\rm CO}=0.8$ for starbursts such as CRLE, AzTEC-3, local starbursts and high-$z$ sub-millimeter galaxies (SMGs). Fixed gas depletion time (corresponding to fixed star formation efficiency) lines are shown for characteristic time-scales spanning from 10 Myr to 5 Gyr (green lines).
}
\label{fig:SF_plot}
\end{figure*}

To study the star formation efficiency in HZ10 and LBG-1, first we directly compare the FIR to CO luminosity (Figure~\ref{fig:SF_plot}) relative to expectations based on previous determinations of the star formation law in the local and high redshift Universe \citep{CarilliWalter}. 
\cite{Daddi10b} and \cite{Genzel10} measured a relationship between the CO luminosity and the SFR (or IR luminosity) for lower redshift Main Sequence disk galaxies which is indicative of an underlying ``star formation law", and found broad agreement within the significant scatter of the observed correlation.
We here aim to investigate its evolution toward higher redshift. The relationship by  \cite{Daddi10b} would predict total IR luminosities of $(2.3\pm0.5)\times10^{12}\,L_\odot$ for HZ10 and $<2.7\times10^{11}\,L_\odot$ for LBG-1, respectively, based on the CO luminosity which are compatible with our direct FIR luminosity estimates.
%On the other hand, the best fit relation for starburst galaxies derived by \cite{CarilliWalter} would over-predict the IR luminosity in HZ10 ($(3.9\pm0.8)\times10^{12}\,L_\odot$) relative to our independent estimate. 
Therefore, we find no evidence for an evolution in the star formation law for Main Sequence galaxies all the way up to $z\sim6$, although larger samples are necessary to statistically assess this finding.

 Adopting our best  estimate of the gas mass and the star formation rate in HZ10 (Table~\ref{table_derived}) yields a gas depletion timescale (the inverse of the star formation efficiency) of $960^{+1200}_{-470}\,$Myr,\footnote{ Uncertainties were propagated from both the star formation rate and the gas mass estimates.} which is significantly longer than what is commonly measured in local and high redshift starburst galaxies ($\lesssim 100\,$Myr; e.g., \citealt{CarilliWalter}).
HZ10 therefore appears to be very rich in molecular gas, and the efficiency of star formation appears compatible with what is commonly observed in lower redshift, disk-like, main-sequence galaxies ($0.5-2\,$Gyr; e.g., \citealt{Leroy13,Tacconi13,Tacconi18}).

The IR luminosity  of LBG-1 is at least a factor of 2 higher than estimates based on its CO luminosity and the star formation law,  suggesting that the CO luminosity in LBG-1 is lower than in  lower redshift galaxies with comparable SFR (Figure~\ref{fig:SF_plot}). If we adopt our best estimates for the SFR in LBG-1 based on the inferred IR luminosity, we can obtain estimates of the gas depletion timescale for the gas masses derived from the long-wavelength dust method ($\sim500^{+500}_{-250}\,$Myr), from the dynamical mass constraints ($\sim280^{+420}_{-180}\,$Myr), and  CO upper limit  $\lesssim2.8$ Gyr  with 99.7\% probability when assuming $\alpha_{\rm CO}=4.5$.

We can also use our estimates for the gas reservoir physical sizes derived from the [C{\sc ii}] line to compare the gas surface density to the star formation rate density probed by the dust continuum flux and size (Figure~\ref{fig:SF_plot}), probing the physical drivers of star formation more directly, i.e., the Kennicutt-Schmidt law \citep{Kennicutt1998,Leroy08,Bigiel08,Daddi10b,Bigiel2011,Schruba11,KennicuttEvans,Leroy13}. 
Specifically, we include in this comparison both Main Sequence galaxies at $z\sim1-3$ \citep{Tacconi13,Daddi10b} and  intensely star forming sub-millimeter galaxies (SMGs, \citealt{Bouche07,Bothwell10}). In particular, we focus our comparison on CRLE and AzTEC-3, two hyper-luminous DSFGs at $z>5$ which are located in  physical proximity of HZ10 and LBG-1, respectively. Based on their global gas masses and SFR, the gas depletion timescales for CRLE and AzTEC-3 are $\sim45-50\,$Myr \citep{Riechers10a,Riechers14,CRLE}, i.e., an order of magnitude shorter than we observe in HZ10 and LBG-1. 
 We    divide each of the SFR and gas masses by the area within the FWHM of the best-fit elliptical Gaussian source model (e.g., \citealt{deBreuck14,Spilker16,Litke19}), uniformly for our sample of two ``normal" and two starburst galaxies at $z=5-6$. Following \cite{Riechers14}, \cite{Spilker16}, and \cite{Hodge16}, we use the dust continuum sizes at $158\,\mu$m as representative of the extent of the star-forming region, since the emission at such short wavelength is dominated by the actively star-forming region. We follow \cite{deBreuck14} and \cite{Litke19} in using the [C{\sc ii}] line emission size as a proxy for the extent of the gas reservoir\footnote{\cite{Spilker16} and \cite{Calistro18}, among others, clarified the importance of not inferring the gas reservoir extent from dust continuum sizes, in this context.}. 
Due to the  compactness of the star formation in AzTEC-3 and CRLE, the local depletion timescale characterizing the ratio of gas and SFR surface densities are as short as $\sim10-30$ Myr, while our estimates for HZ10 and LBG-1 are $\sim1$ Gyr and $\lesssim300$ Myr, respectively (Figure~\ref{fig:SF_plot}). Therefore, the physical efficiency, in terms of surface densities, may potentially differ by up to two orders of magnitude already among these galaxies at $z>5$.
The comparison shown by Figure~\ref{fig:SF_plot} shows that, while AzTEC-3 and CRLE have high star formation efficiency compatible with other high redshift starbursts, HZ10 (and to a lesser degree LBG-1) appear to exhibit the lower efficiencies, longer depletion times typically observed in Main Sequence disks at lower redshift. 
Although the depletion time measurement in HZ10 is incompatible with that in  starbursts (e.g., \citealt{Silverman15,Silverman18}), the  systematic uncertainty implies compatibility with both the efficiency in $z\sim0$ disk galaxies (e.g., \citealt{Leroy13}) but also the potentially higher efficiency suggested for Main Sequence galaxies by \cite{Tacconi13}, \cite{Genzel15}, \cite{Scoville16,Scoville17}, and \cite{Tacconi18}.

\begin{figure}[htb]
\centering{
 \includegraphics[width=.5\textwidth]{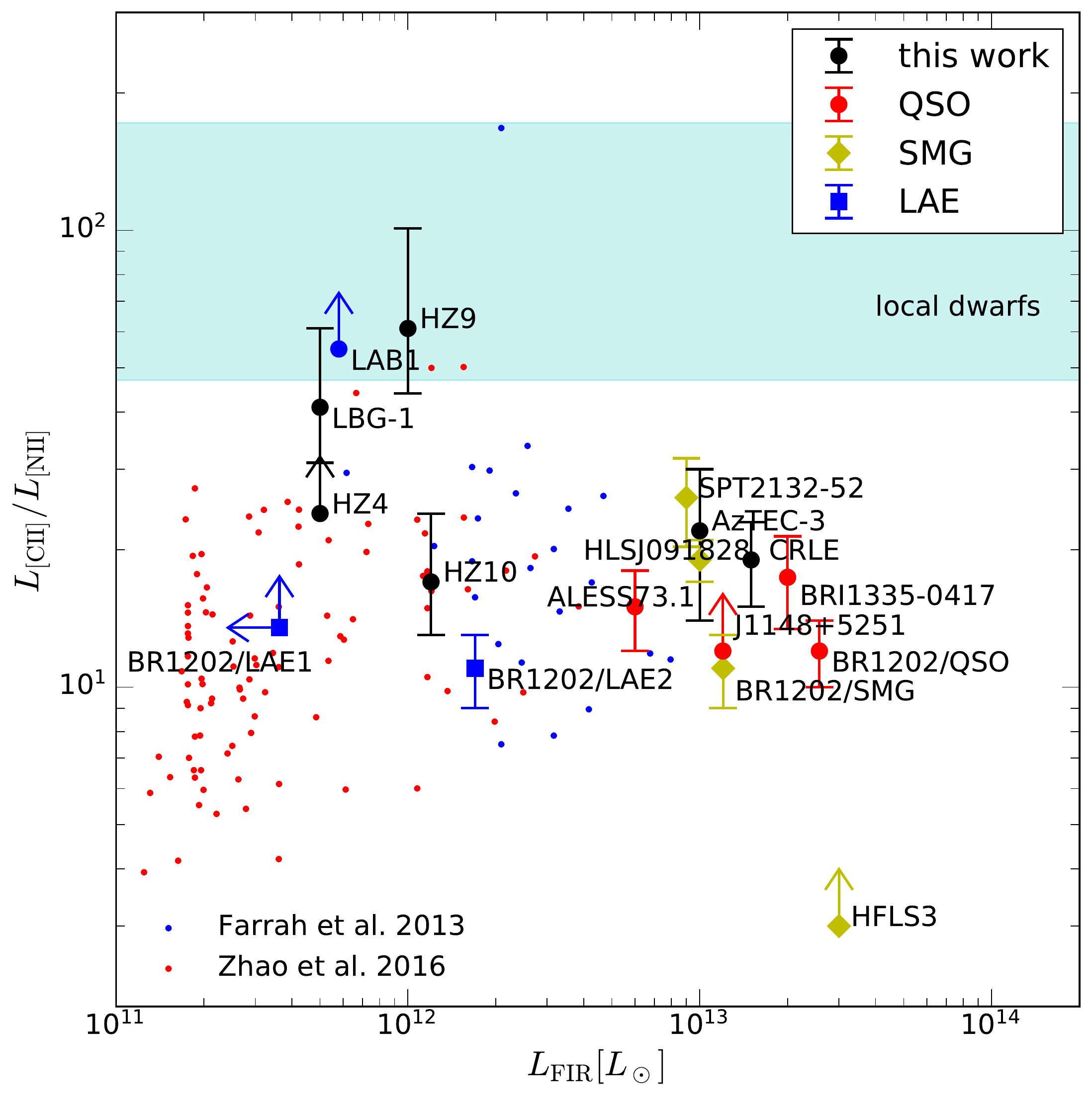} 
}
\caption{[C{\sc ii}]/[N{\sc ii}] line luminosity ratios observed in high redshift galaxies to-date as a function of their FIR luminosity (\citealt{CarilliWalter,deBreuck14,Bethermin16, Rawle14, Combes12, Carilli13,RiechersHFLS3,Riechers14,C15,WaggBRI,Umehata2017,LuBRI,CRLE}). For comparison we also show a sample of local ULIRGs from \cite{F13} (using the [N{\sc ii}] $122\,\mu$m line) and LIRGs with [N{\sc ii}] (using the [N{\sc ii}] $205\,\mu$m line) from \cite{Z16} and [C{\sc ii}] from \cite{DS}. The range of ratios in dwarfs \citep{Cormier15} (using the [N{\sc ii}] $122\,\mu$m line) is shown as a cyan band. The [N{\sc ii}] $122\,\mu$m line measured in the indicated local samples was converted to a [N{\sc ii}] $205\,\mu$m luminosity assuming a ratio of 1/2.5, estimated by \cite{HerCam16}. The abscissa in the local samples are defined as total IR luminosity; no attempt was made to convert to a common FIR luminosity scale because it does not affect our interpretation.}
\label{fig:ratio_plot}
\end{figure}

\subsection{[N{\sc ii}] measurements}
 To complement our view of the star forming gas in our sample galaxies, we observed the [N{\sc ii}] $205\,\mu$m line emission which is one of the best tracers of the ionized component of the ISM. The [C{\sc ii}]/[N{\sc ii}] line ratio is the tool of choice to determine the fraction of [C{\sc ii}] emission coming from ionized gas (e.g., \citealt{Oberst2006,DecarliNII,Pavesi16}). This quantity is itself a probe of the physical conditions of the gas which is directly exposed to recent star formation, and may be a probe of the radiation intensity, hardness and hence, indirectly, of the gas-phase metallicity. Metallicity directly affects our interpretation of  CO observations (since it is one of the main drivers of the variation in $\alpha_{\rm CO}$), and it offers unique insights into the balance between fresh gas inflow and ISM enrichment through previous star formation. The analysis in this section expands on previous analysis presented by \cite{Pavesi16}.

We tentatively detect [N{\sc ii}] $205\,\mu$m emission toward LBG-1 (at $\sim3.4\sigma$), and HZ9 (at $\sim3.1\sigma$), we confidently detect it from HZ10 (at $\gtrsim6\sigma$), and we provide a constraining upper limit for HZ4 (Figures~\ref{fig:mom_cont} \& \ref{fig:spectra}).
 We confirm previous results on LBG-1 and  HZ10 \citep{Pavesi16} by achieving a higher signal-to-noise ratio.
We measure [N{\sc ii}] and [C{\sc ii}] line properties using aperture spectra consistently for the whole sample (Table~\ref{table_lines}, see footnote~\ref{aper_foot}).
The [N{\sc ii}] emission from HZ10 appears extended, at limited significance ($\sim2-3\sigma$, Figure~\ref{fig:mom_cont}). In order to measure the  [N{\sc ii}] emission size, we fit a circular Gaussian model to the integrated [N{\sc ii}] line visibilities from HZ10,  using {\sc casa} {\sc uvmodelfit}.  We estimate a deconvolved [N{\sc ii}] spatial FWHM size of $1^{\prime\prime}.71\pm0^{\prime\prime}.25$ for HZ10, formally corresponding to $10\pm2$ kpc. We use the same technique to measure an effective circular [C{\sc ii}]  size of $0^{\prime\prime}.61\pm0^{\prime\prime}.04$, corresponding to $3.6\pm0.2$ kpc, which is compatible with our more sophisticated {\sc uv} plane modeling (Appendix B). The [N{\sc ii}] line emission could be marginally more extended than the [C{\sc ii}] emission,  but higher resolution and higher signal to noise [N{\sc ii}] observations are necessary to confirm this tentative finding. In particular, a manual inspection of the {\sc uv}-radial profile of the [N{\sc ii}] line visibilities appears compatible with the size of the [C{\sc ii}] emission within the relative uncertainties.

We do not detect spatial offsets between the [N{\sc ii}] and [C{\sc ii}] line emission in LBG-1, HZ9 and HZ10. Although we now confirm that the [N{\sc ii}] line emission in HZ10 comes from the full [C{\sc ii}] velocity range, the comparison of the [N{\sc ii}] and [C{\sc ii}] spectra (Figure~\ref{fig:spectra}) suggests a possible differential intensity ratio, with stronger [N{\sc ii}] intensity coming from the red part of the emission which dominated the lower signal-to-noise detection found by \citealt{Pavesi16}.
The [N{\sc ii}]  line velocity width appears narrower than [C{\sc ii}] toward LBG-1 and HZ9 at the current sensitivity of our measurements, although the limited signal-to-noise does not allow for a reliable measurement of the line widths .

To enable a more comprehensive study, we here update our measurement of the [C{\sc ii}]/[N{\sc ii}] line ratio for HZ10 and LBG-1 \citep{Pavesi16}, and we expand the sample to include HZ9 and HZ4 (Table~\ref{table_lines}, Figure~\ref{fig:ratio_plot}).
We confirm the relatively low line ratio for HZ10, which is compatible with most local and high redshift active star-forming galaxies \citep{Pavesi16}. On the other hand, we  find substantially higher ratios for HZ9, LBG-1 and HZ4, which are only compatible with the ratio observed in local dwarf galaxies (Figure~\ref{fig:ratio_plot}) and provides further evidence for the diversity of conditions at $z>5$ already present in this small sample.

As previously described by \cite{Pavesi16}, a high [C{\sc ii}]/[N{\sc ii}] line ratio may be expected in the case of very high gas density, or of high intensity and hardness of the radiation field. The latter explanation appears consistent with observations of a high line ratio in local dwarf galaxies by \cite{Cormier15}, which may be interpreted as the consequence of high radiation intensity and hardness due to low metallicity conditions. A high intensity and hardness radiation field is expected to induce higher ionization states in the ionized gas. This implies weak [N{\sc ii}] and  [C{\sc ii}]  emission from ionized gas because nitrogen, carbon  and oxygen are expected to be in a higher ionization state. This prediction is testable by observing strong [N{\sc iii}] $57\,\mu$m and [O{\sc iii}] $52\,\mu$m or $88\,\mu$m emission lines.
 \cite{Croxall17} recently confirmed this interpretation using observations of local star-forming galaxies from the KINGFISH sample. They conclusively reported a strong correlation of the  [C{\sc ii}]/[N{\sc ii}] line ratio with gas-phase metallicity, concluding that metallicity appears to be the main driver of this line ratio.
Therefore, in analogy to the case of local dwarf galaxies, we interpret our high line ratio measurements and limits for HZ9, LBG-1 and HZ4 as indicative of low gas (and stellar) metallicity, relative to $z<5$ galaxies of comparable masses ($\sim10^{10}\,M_\odot$). On the other hand, the lower line ratio observed in HZ10 suggests higher metallicity in this galaxy, confirming the inference based on high dust and CO emission, and suggesting a particularly ``mature, normal" galaxy at the same epoch  (see also discussion by \citealt{Faisst17}).
We note that an alternative interpretation for the low [N{\sc ii}] luminosity in our high redshift sample may invoke lower nitrogen abundance, relative to carbon. While this abundance ratio change may also be a consequence of low metallicity conditions due to the secondary nature of nitrogen, the carbon abundance dependence on gas phase metallicity is not well constrained.
We follow \cite{Tanio17} and assume a line ratio of $3\pm0.5$ in the ionized gas to infer fractions of [C{\sc ii}] coming from PDRs of $83\%\pm6\%$ for HZ10, $96\%\pm2\%$ for HZ9, $93\%\pm3\%$ for LBG-1 and $>86$\% for HZ4.

\section{Discussion}

The detection of bright CO emission from HZ10 represents the highest redshift CO detection from a ``normal", Main Sequence galaxy to date (the next highest redshift CO line from an unlensed Main Sequence galaxy was serendipitously detected by \citealt{Avani,Avani19} at $z\sim3.2$). 
We note that HZ10 appears to have a very high gas fraction based on the measured CO luminosity ($M_{\rm gas}>M_{\rm stars}$ with high confidence and likely $M_{\rm gas}\sim4-5 \times M_{\rm stars}$, Figure~\ref{fig:gas_frac}). Such high gas fractions may be expected at $z>5$ based on the extrapolation of observed trends (e.g., \citealt{Magdis2012,Genzel15,Tacconi18}). A high gas fraction may also potentially be connected with the possibility of a galaxy merger in HZ10 since merging galaxies have been found to potentially show enhanced gas fractions (e.g., \citealt{Pavesi_COLDz}). Assuming $\alpha_{\rm CO}\sim4.5$  would imply a $\gtrsim4$ times lower gas fraction for LBG-1, potentially suggesting significant scatter within the general population. However, the low CO luminosity in LBG-1 is likely to be due to low metallicity, as suggested by the faint [N{\sc ii}] emission, and the gas fraction may therefore be substantially higher in practice.

\cite{ZavalaNature} recently reported observations of CO and [C{\sc ii}] emission lines from the strongly lensed galaxy G09 83808 at $z\sim6$, presenting analogies with HZ10. Although the inferred CO(1--0) luminosity of G09 83808 is approximately three times lower than HZ10, the dust continuum emission is at least twice as bright at rest-frame 158 $\mu$m, indicating a significantly higher star formation efficiency than found in HZ10. Therefore, while G09 83808 appears to have only few times higher star formation rate than HZ10, its star formation  properties resemble starbursts such as CRLE and AzTEC-3, while HZ10 is more gas rich and exhibits star forming conditions compatible with lower redshift Main Sequence, disk galaxies.
This finding is in agreement with the ten times higher [C{\sc ii}]/FIR ratio in HZ10 relative to G09 83808. 
This ratio is a probe of the local physical density of star formation and is inversely proportional to the starburst intensity. Based on PDR models, a fixed PDR gas density implies that the far UV (FUV) field intensity (G$_0$) scales inversely with [C{\sc ii}]/FIR  (to a power of $\sim1-1.2$) \citep{Wolfire90,Kaufman99,Stacey10}.  This scaling implies that the FUV intensity in G09 83808 may be $\sim10-15$ times higher than in HZ10, confirming that HZ10 may be forming stars in a much less intense environment.

The finding of significant dust and, especially, CO emission from HZ10 suggests that a fraction of ``normal" galaxies (not extreme starbursts) at $z>5$ may be rich in molecular gas and significantly metal-enriched, in contrast to some previous indications (e.g., \citealt{Tan13,Tan14}).
This finding is in agreement with the recent measurement of a high volume density of CO-selected galaxies at $z>5$ by the CO Luminosity Density at High-z (COLDz) project \citep{Pavesi_COLDz,Riechers_COLDz}. Although the galaxies selected by COLDz at $z>5$ are bright starbursts, their volume density is significantly higher than predicted by current models \citep{Riechers_COLDz}. If HZ10 had been located within the COLDz field of view it would have been selected by the blind line search based on the survey detection limit \citep{Pavesi_COLDz}, therefore placing an upper limit to the volume density of evolved, gas-rich ``normal" galaxies at $z>5$ with CO luminosity greater than HZ10 of $\lesssim5\times10^{-5}\,$Mpc$^{-3}$ \citep{Riechers_COLDz}. %5e-5 Mpc^-3

HZ10 is believed to reside in a galaxy overdensity at $z\sim5.7$, potentially indicating a protocluster environment \citep{CRLE}. In particular, the presence of the bright hyper-starburst CRLE only $\sim70\,$kpc away constitutes evidence for a possible physical association. This association with the massive, dusty galaxy CRLE and the protocluster may be related to the advanced evolutionary stage of HZ10. If this connection were confirmed, it would point to a more rapid evolution for galaxies in higher density environments (e.g., \citealt{Chiang2017}).

The PHIBBS survey has measured star formation efficiency and gas fractions for lower redshift  Main Sequence galaxies (up to $z\sim2-3$; \citealt{Tacconi13,Genzel15,Tacconi18}). Based on the extrapolation of the latest measured trends reported by \cite{Tacconi18}, combining the PHIBBS CO measurements with the dust-based estimates by \cite{Scoville16,Scoville17} we can estimate the average gas fractions and depletion times expected for Main Sequence galaxies such as HZ10 at $z\sim5.7$. We derive an approximate gas depletion timescale of $\sim400$ Myr, which is compatible with our estimate for HZ10 within $1\sigma$. The molecular gas fraction predicted by the fitting formula suggested by \cite{Tacconi18} is $M_{\rm gas}\sim M_{\rm stars}$\footnote{The quadratic fitting formula predicts, perhaps artificially, a turnover of the trend  at $z\sim3.5$}, which is lower than observed in HZ10. Our observations therefore suggest that the increase in molecular gas fraction with redshift may continue beyond $z\sim3$, although with  limited statistical power due to the small sample size. In summary, HZ10 shows the characteristic properties of lower redshift Main Sequence galaxies, all the way back to the first billion years of cosmic time.

\cite{Vallini18} presented some of the latest models of the CO line emission from ``normal" galaxies at $z>5$. They modeled the radiative transfer affecting CO emission from a clumpy molecular medium in a $M_{\rm stars}\sim10^{10}\,M_\odot$ Main Sequence galaxy at $z\sim6$. Although their model galaxy is characterized by sub-solar (0.5$Z_\odot$) metallicity they predict a low effective CO conversion factor of $\alpha_{\rm CO}\sim1.5$ due to the dominant effect of warmer gas, high turbulence and high gas surface density \citep{Vallini18}. While such a low $\alpha_{\rm CO}$  may be allowed for HZ10, it is ruled out for the more typical LBG-1 if the gas mass is predominantly molecular. In addition, the predicted CO luminosity for the ``typical" model galaxy is $\sim20$ times lower than observed in HZ10, suggesting that the molecular gas mass may be significantly underestimated. Therefore, HZ10 may be more mature and may therefore not be analogous to the model galaxy, but rather to the lower redshift Main Sequence galaxies observed at $z\sim2-3$. Although our constraints for the CO luminosity in LBG-1 are compatible with the model predictions, the higher dynamical mass estimates suggest higher gas masses for LBG-1 than the molecular mass predicted by the models. A possible interpretation of this result may invoke a significant fraction of gas in the atomic phase, which may dominate the total gas mass in such ``typical", massive galaxies. Based on the [C{\sc ii}] luminosity in LBG-1 we can derive an estimate of the atomic PDR mass of $\sim2-5\times10^9\,M_\odot$ (following \citealt{Stacey91}), which may be comparable to the molecular gas mass for low $\alpha_{\rm CO}$, but it is unlikely to provide the total gas mass inferred from our dynamical mass estimate.

The measurement of [C{\sc ii}] and dust continuum emission from the first sample of ``normal", rest-UV selected galaxies revealed a variety of star-forming  conditions \citep{C15}. 
The finding of bright CO line emission from HZ10 and faint emission from LBG-1 is in agreement with the interpretation of a range of metallicities and dust-to-gas ratios being the main contributors to the variation within the sample \citep{C15}. This interpretation is strongly supported by the significant difference in [C{\sc ii}]/[N{\sc ii}] ratios between HZ10 and LBG-1 already noted by \cite{Pavesi16}. 
Faint [N{\sc ii}] emission relative to [C{\sc ii}] directly implies (with the possible caveat of differences in the C/N abundance ratio) low contribution of the ionized gas to the [C{\sc ii}] emission, which may therefore be predominantly due to emission from neutral PDRs. The simplest interpretation for faint [N{\sc ii}] emission suggests higher ionization conditions in the ionized gas, predicting bright [N{\sc iii}] and [O{\sc iii}] emission, instead. This interpretation would suggest that intensity and, especially, hardness of the radiation field may be the most relevant physical parameter affecting this line ratio.
Recent detections of bright [O{\sc iii}] 88 $\mu$m line emission at high redshift support this interpretation and suggest that [O{\sc iii}] may be even brighter than [C{\sc ii}] in ``normal" galaxies at very high redshift (e.g., \citealt{Inoue2016,Laporte2017,Carniani17,Marrone2018,Hashimoto2018_three,
Hashimoto2018_nature,Hashimoto2018_quasar,Tamura2018}), as typically observed in local dwarfs \citep{Cormier15}.
Furthermore, recent optical studies of LBGs and LAEs have also found increasing [O{\sc iii}]$\lambda5008$  brightness at high redshift together with high sSFR and low metallicity (e.g., \citealt{Strom17b,Strom17a}). The metallicity dependence may also be responsible for the downturn due to reduced oxygen abundance at even lower metallicity \citep{Harikane18}.

\cite{Faisst17} explore the level of maturity, stellar population properties and dust attenuation in $z=5-6$ ``normal" galaxies through the IRX/$\beta_{\rm UV}$ diagnostic plane. 
While IRX, defined as the ratio $L_{\rm IR}/L_{\rm UV}$, represents the prevalence of dust-obscured star-formation, $\beta_{\rm UV}$ is the power-law slope of the UV emission, which bears the imprint of dust reddening. A correlation between these quantities was observed to hold for local starburst galaxies, and approximately holding up to high redshift (e.g., \citealt{Meurer99, Reddy06,Reddy10,Reddy18, Bouwens16}), however variations may be expected due to varying dust properties, star-formation geometry and stellar population ages (e.g., \citealt{Faisst17, Narayanan18}).
 These diagnostics suggest that HZ10 may resemble dusty star-forming galaxies, with elevated IR to UV luminosity ratio, intriguingly sharing similarities to lower redshift IR-selected galaxies (e.g., \citealt{Casey2014b}). However, HZ10 was selected through the LBG and LAE techniques at $z\sim5.7$ and appears ``typical" based on its UV emission. In particular, HZ10 lies within the scatter of the Main Sequence at this redshift (e.g., \citealt{Speagle14,C15,Barisic17,Faisst17}). \cite{Faisst17} also interpret the observed properties of LBG-1 as being consistent with lower dust and metal abundances, likely connected to young stellar populations. The IRX/$\beta_{\rm UV}$ diagnostic, however, would suggest that HZ4, and especially HZ9, may be more dusty than LBG-1 since they lie on or above the local \cite{Meurer99} relation \citep{Faisst17}. However, the measured [C{\sc ii}]/[N{\sc ii}] ratios for HZ4 and HZ9 are compatible with that in LBG-1 and significantly higher than the ratio in HZ10 (Figure~\ref{fig:ratio_plot}). The intriguing finding of faint [N{\sc ii}] emission together with relatively bright dust continuum in HZ9 therefore suggests the presence of additional variables controlling the relationship between the level of dust obscuration and the metallicity (or age of the most recent stellar population) which may be critical to diagnose the interplay between gas inflows, outflows and star formation. An important next step would require measuring the CO line luminosity from HZ9 as well as achieving a detection in LBG-1. In case of relatively bright CO emission (e.g., in relation to its FIR luminosity) from HZ9, the high [C{\sc ii}]/[N{\sc ii}] line ratio would not be explained by the analogy to local dwarf galaxies and would point to previously unexplored star formation conditions. However, faint CO line emission from HZ9 would either suggest variations in the dust SED shape  or would intriguingly suggest the possibility of significant dust-obscured star formation even in more ``typical", lower metallicity, younger high redshift galaxies. The ratio of our continuum measurements tentatively suggests higher dust temperatures in HZ9 than in HZ10. If correct, this might imply that the moderate IR luminosity in HZ9 may be due to higher temperatures, perhaps associated with higher radiation intensity, rather than a high dust content \citep{Faisst17}.  \cite{Bethermin15} already presented evidence in favor of such rising radiation field intensity and dust temperatures toward higher redshift, and showed that these may be a direct consequence of decreasing metallicity. \cite{Ferrara_cold_dust} suggested that galaxies at $z>5$ may be FIR-faint due to colder dust than ``normal" due to the very high molecular gas fraction. Their prediction of bright CO emission, specifically from galaxies with low IRX, may be in conflict with our deep upper limits on the CO luminosity from LBG-1. However, this effect may link the high molecular gas mass fraction in HZ10 to the tentatively lower dust temperature we observe in this galaxy relative to the rest of the sample \citep{Ferrara_cold_dust}.

The faint [N{\sc ii}] emission from HZ4 and HZ9, together with significant dust-obscured star formation, may be analogous to the properties observed in the eastern component of SPT0311-58 \citep{Marrone2018}. This galaxy at $z=6.90$ was shown to display high [O{\sc iii}] 88$\mu$m luminosity ($\sim2\times$ its [C{\sc ii}] luminosity) while being characterized by  very high dust-obscured star formation (at the level observed in HZ9 and HZ10). Similarly, the bright [O{\sc iii}] emitters studied by \cite{Hashimoto2018_three} and \cite{Tamura2018} at $z>7$ which also show significant dust emission may be somewhat analogous to the case we observe in HZ9, i.e., high intensity and hardness of the radiation causing a higher ionization state in the ionized ISM while showing significant dust-obscured star formation.
Furthermore, a comparison of the [O{\sc iii}]/[C{\sc ii}] luminosity in two quasars at $z\sim6$ suggests that this line ratio may strongly correlate with dust temperatures \citep{Hashimoto2018_quasar}, supporting our interpretation of higher dust temperatures in [N{\sc ii}]-faint galaxies.
%The dust continuum at rest-frame 158 $\mu$m from SPT0311-58 E is similar to that from HZ10 and $\sim2\times$ lower than from HZ9, but the [C{\sc ii}] luminosity is approximately equal, 
We therefore suggest that a higher dust temperature may drive the observed FIR luminosity in such galaxies, perhaps due to a significant contribution from dust in the ionized regions \citep{Faisst17}. 

In order to assess how common the different star forming conditions observed in LBG-1, HZ9 and HZ10 are, larger samples of ``normal" galaxies at $z=5-6$ need to be studied. The  ALMA Large Program to Investigate [C{\sc ii}] at Early Times (ALPINE)\footnote{https://cesam.lam.fr/a2c2s/index.php}, is now observing the [C{\sc ii}] and dust emission from large samples of typical galaxies at $4<z<6$ over a wide range of stellar mass and star formation rate.
While the brightness of [C{\sc ii}] and dust continuum, and their relation to the ultra-violet flux, provide a wealth of information (e.g., distinguishing LBG-1 from HZ10-type conditions), our analysis shows that relevant residual degrees of freedom are unconstrained unless either CO or a tracer of the ionized gas (such as [N{\sc ii}], [N{\sc iii}] or [O{\sc iii}]) is measured in addition to [C{\sc ii}] (to distinguish HZ10 from HZ9-type conditions) possibly due to metallicity, and/or dust temperature variations. Furthermore, resolved observations for larger samples of galaxies are necessary because accurate dynamical masses may be the best way to constrain the gas mass and, hence, to directly infer the $\alpha_{\rm CO}$ conversion factor.

\section{Conclusions}
We have presented measurements of CO(2--1) line emission from two ``normal" Lyman Break Galaxies galaxies, at the end of the ``Epoch of Reionization", achieving the highest redshift low-{\em J} CO detection from a Main Sequence galaxy to date. We have found large variation in the CO line luminosity between the two targeted sources which may not be completely accounted for by SFR differences (the CO luminosity ratio is $\gtrsim$6.5 while the SFR ratio is $\sim3$).
While this difference in CO luminosity may suggest variations in star formation efficiency, it appears consistent with our expectation of lower gas metallicity and dust abundance strongly affecting the CO abundance.
We infer a large molecular gas reservoir in at least one of the  sources, suggesting low efficiency star formation with gas depletion time $\sim$1 Gyr already at $z\sim6$, analogous to what is commonly observed in lower redshift disk galaxies. This low efficiency contrasts to what is typically observed in $z>5$ starbursts and provides the first evidence of such ``Main sequence" star-forming conditions at $z>3$.
We also find evidence for a continuously rising gas fraction up to $z\sim6$, although our sample may suggest either significant scatter or systematic variations in the $\alpha_{\rm CO}$ conversion factor.

By observing the largest sample of ``normal" galaxies at $z>5$ in [N{\sc ii}] 205$\,\mu$m emission to date, we find a general trend of increasing [C{\sc ii}]/[N{\sc ii}] ratios with lower IR luminosity; consistent with what was previously reported by \cite{Pavesi16}. Our findings support an interpretation where low gas and stellar metallicity raise the ionization state of carbon and nitrogen in the ionized gas. This interpretation suggest that the large majority of [C{\sc ii}] emission from most ``normal" galaxies at $z>5$ may emerge from the neutral gas phase. We also find a high [C{\sc ii}]/[N{\sc ii}] ratio in our sample with moderate IR luminosity, suggesting either significant dust temperature variations affecting the IR luminosity estimate, or the possibility of a young starburst with high radiation intensity and hardness (and potentially low metallicity) together with substantial dust obscuration. 
Our findings imply that a significant fraction of Main Sequence star formation taking place up to $z\sim6$ may resemble the conditions observed in ``normal"  galaxies at lower redshift, suggesting that the efficiency of star formation may only weakly depend on those physical properties which are affected by redshift evolution. In particular, the high inferred gas fractions and the higher merger rates do not appear to significantly affect Main Sequence star formation. Although low metallicity may be common  in the Main Sequence galaxy population at $z>5$, we do not find conclusive evidence for an effect on the star forming conditions, although larger samples and more sensitive observations are needed to study this fainter population.

\smallskip
\textbf{Acknowledgments}
We thank Chelsea Sharon and Avani Gowardhan for useful discussion. 
R.P. and D.R. acknowledge support from the National Science Foundation under grant number
AST-1614213 to Cornell University.
R.P. acknowledges support through award SOSPA3-008 from the NRAO. 
The National Radio Astronomy Observatory is a facility of the National Science Foundation operated under cooperative agreement by Associated Universities, Inc.
This paper makes use of the following ALMA data: ADS/JAO.ALMA\#2015.1.00928.S, 2015.1.00388.S, 2012.1.00523.S, 2011.0.00064.S. ALMA is a partnership of ESO (representing its member states), NSF (USA) and NINS (Japan), together with NRC (Canada), NSC and ASIAA (Taiwan), and KASI (Republic of Korea), in cooperation with the Republic of Chile. The Joint ALMA Observatory is operated by ESO, AUI/NRAO and NAOJ.

\appendix
\section{A. Spectral energy distribution}
\begin{figure*}[htb]
\centering{
 \includegraphics[width=\textwidth]{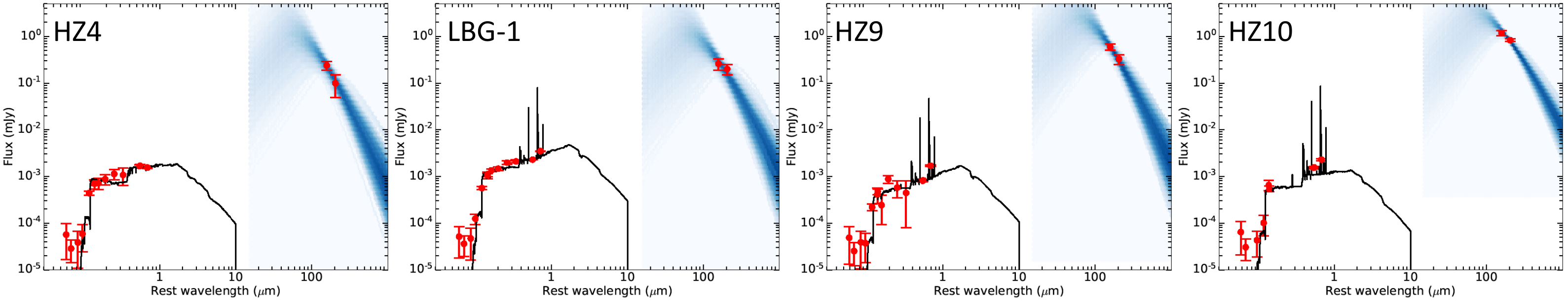}
 }
\caption{Optical to FIR SED for the full galaxy sample. Observed optical to NIR fluxes (from the catalog presented by \citealt{Laigle16}) and FIR fluxes presented here are shown as red markers. Optical-to-NIR fitting to the stellar emission is shown in black \citep{C15}. Modified blackbody fitting to the FIR emission described in the text is shown as a blue shading, with darker shading indicating higher probability areas.
}
\label{fig:full_sed_appendix}
\end{figure*}

Here we present the results of modified blackbody fitting to the available dust continuum observations in the full galaxy sample. The limited sampling of the dust emission is responsible for the high uncertainties on the predicted FIR peak. We illustrate the results of our probabilistic analysis by a gradient of color shading, derived using the results of the MCMC samples, showing the fitting results and higher confidence regions in darker shading (Figure~\ref{fig:full_sed_appendix}).
We also present best fit stellar emission models to the archival optical and NIR observations as templates, previously described by \cite{C15}.

\section{B. Dynamical Modeling}
We have carried out a dynamical modeling analysis directly on the visibilities for the [C{\sc ii}] observations in HZ9 and HZ10 using \texttt{GALARIO} \citep{GALARIO} and \texttt{Multinest} \citep{Multinest},  using the method\footnote{ Python code available at https://github.com/pavesiriccardo/UVmodeldisk} previously described by \cite{CRLE}. While the  [C{\sc ii}] line in HZ9 and HZ10 shows a smooth velocity gradient, the line in LBG-1 shows a more complex morphology and dynamics, with three components, and two separate velocity gradients \citep{Riechers14}. Therefore we do not attempt modeling the emission from LBG-1, as the data are not sufficient to properly constrain such a high-complexity model. Although the [C{\sc ii}] line in HZ10 shows a smooth velocity gradient, the {\em HST} NIR  and  dust continuum images from ALMA suggest the presence of two separate morphological components. These may be associated with either a galaxy merger, or with clumpy gas and stellar distributions, embedded in a rotating disk. The somewhat asymmetric [C{\sc ii}] line profile may also be caused by massive gas clumps, as shown by the simulations of \cite{Daddi10a,Bournaud14,Bournaud15}.

\begin{figure*}[htb]
\centering{
 \includegraphics[width=.45\textwidth]{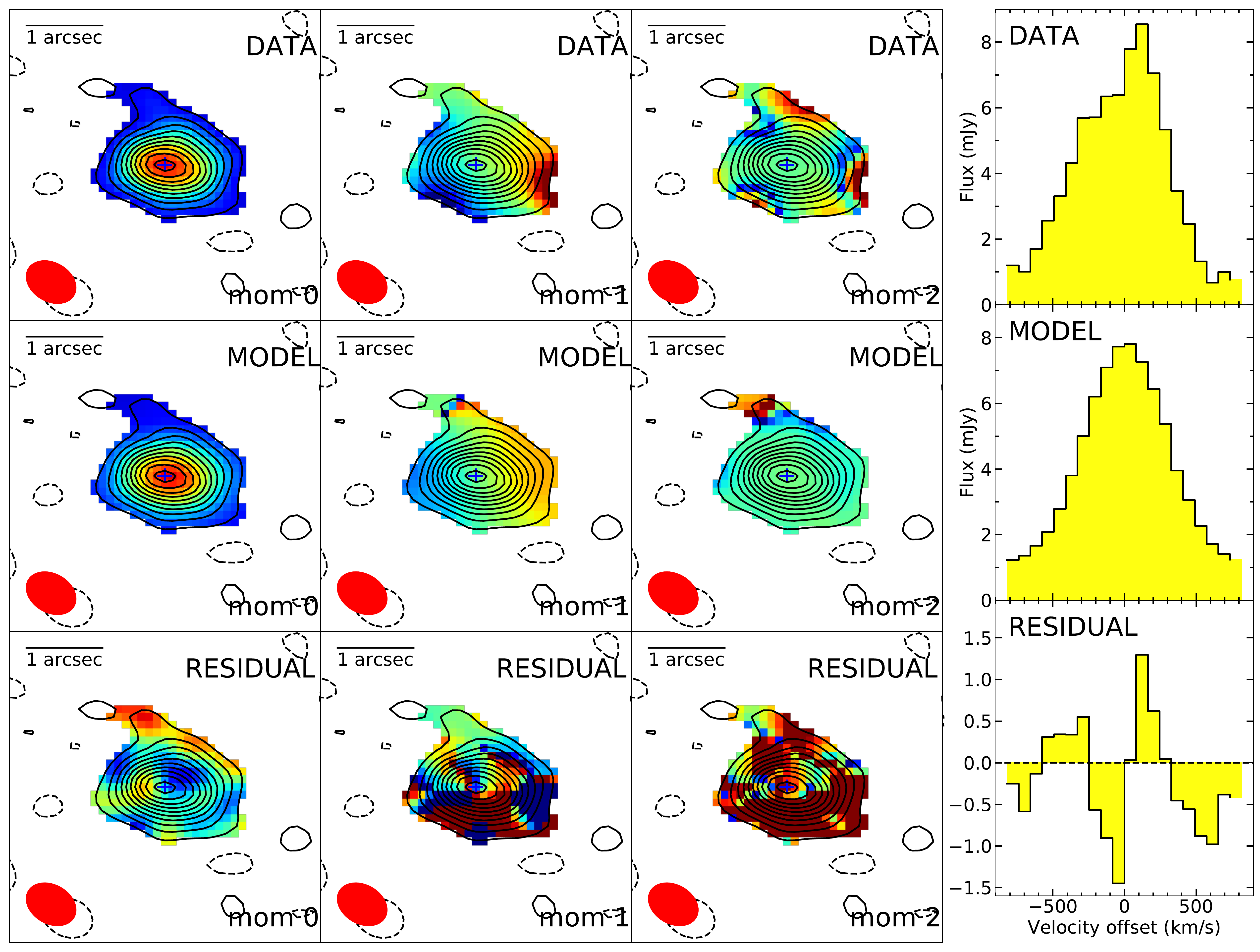} \includegraphics[width=.45\textwidth]{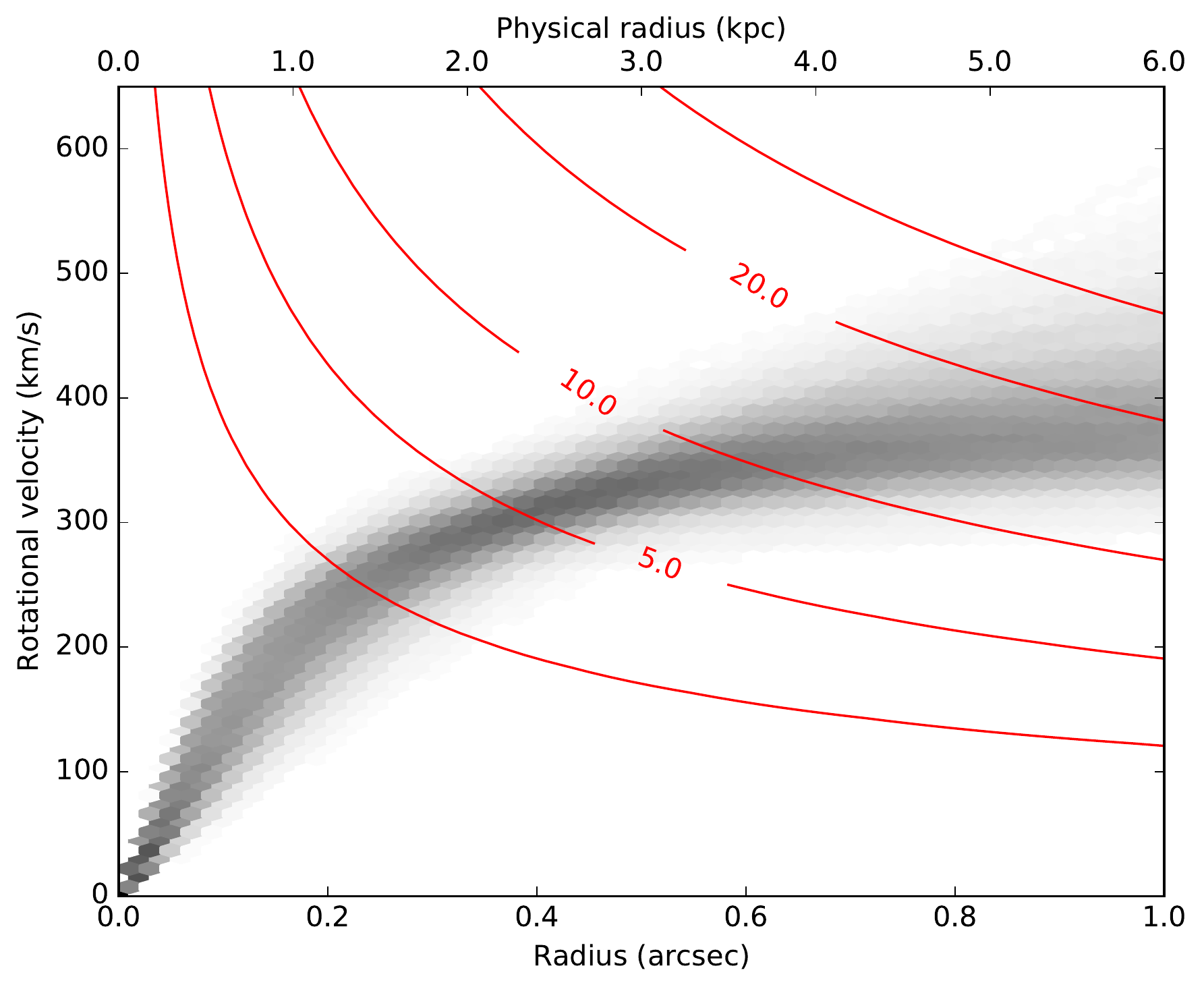}
 }
\caption{Left: Visibility space dynamical modeling results for the [C{\sc ii}] line emission in HZ10. We show the ``natural" weighting line moment 0 (intensity), 1 (velocity) and 2 (dispersion) maps and spectra for the data, the single-disk model corresponding to posterior median parameters, and the visibility residuals. Two-Gaussian components were adopted as model for the continuum. Right: Probabilistic constraints to the rotation curve for a ``tan" model with two disk modeling parameters (maximum velocity and half-velocity radius). The darker shading corresponds to higher probability density, as determined by the MCMC samples. We also show the enclosed dynamical mass in units of 10$^{10}\,M_\odot$ (red curves).}
\label{fig:dyn_mod_HZ10}
\end{figure*}

\begin{figure*}[htb]
\centering{
  \includegraphics[width=.45\textwidth]{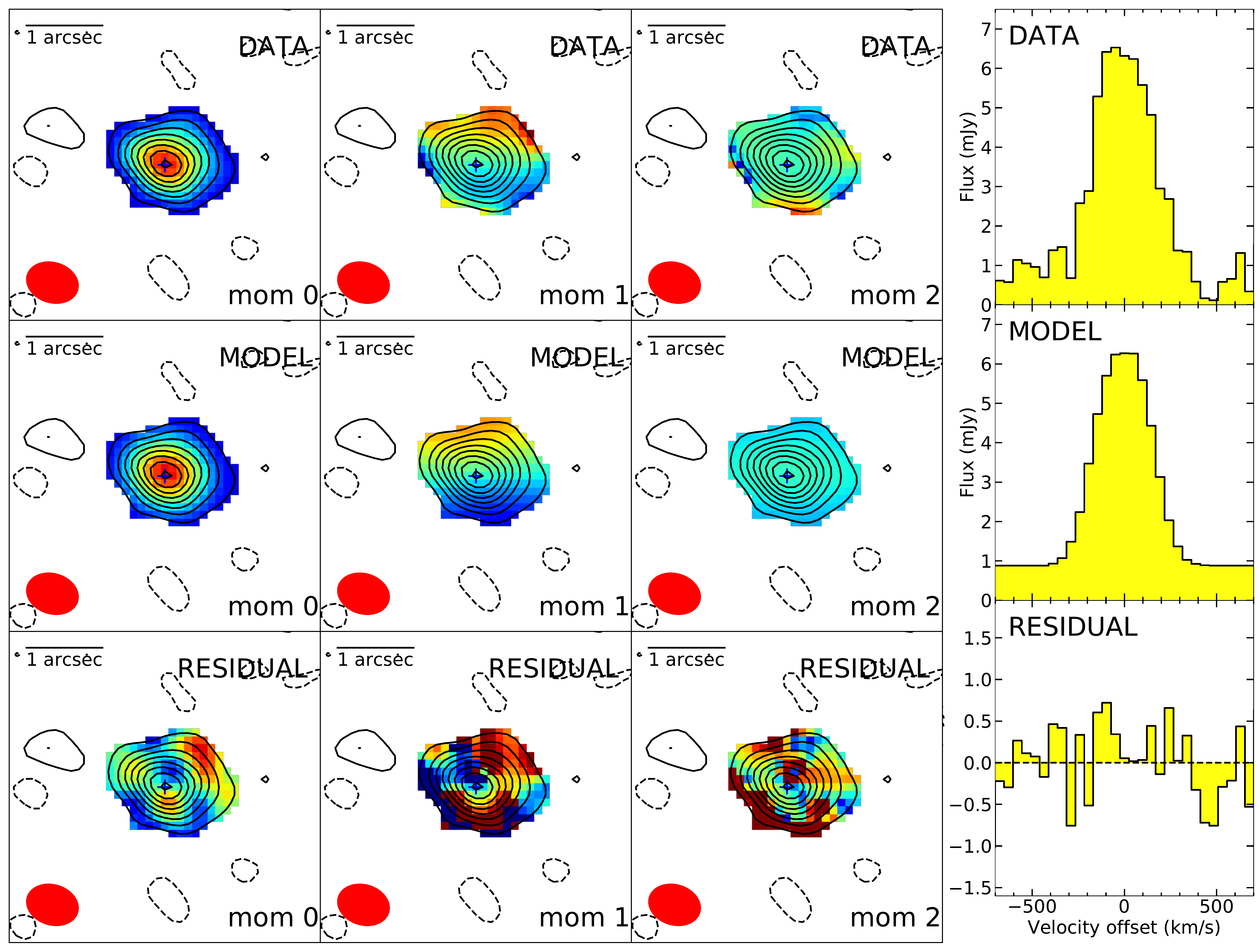} \includegraphics[width=.45\textwidth]{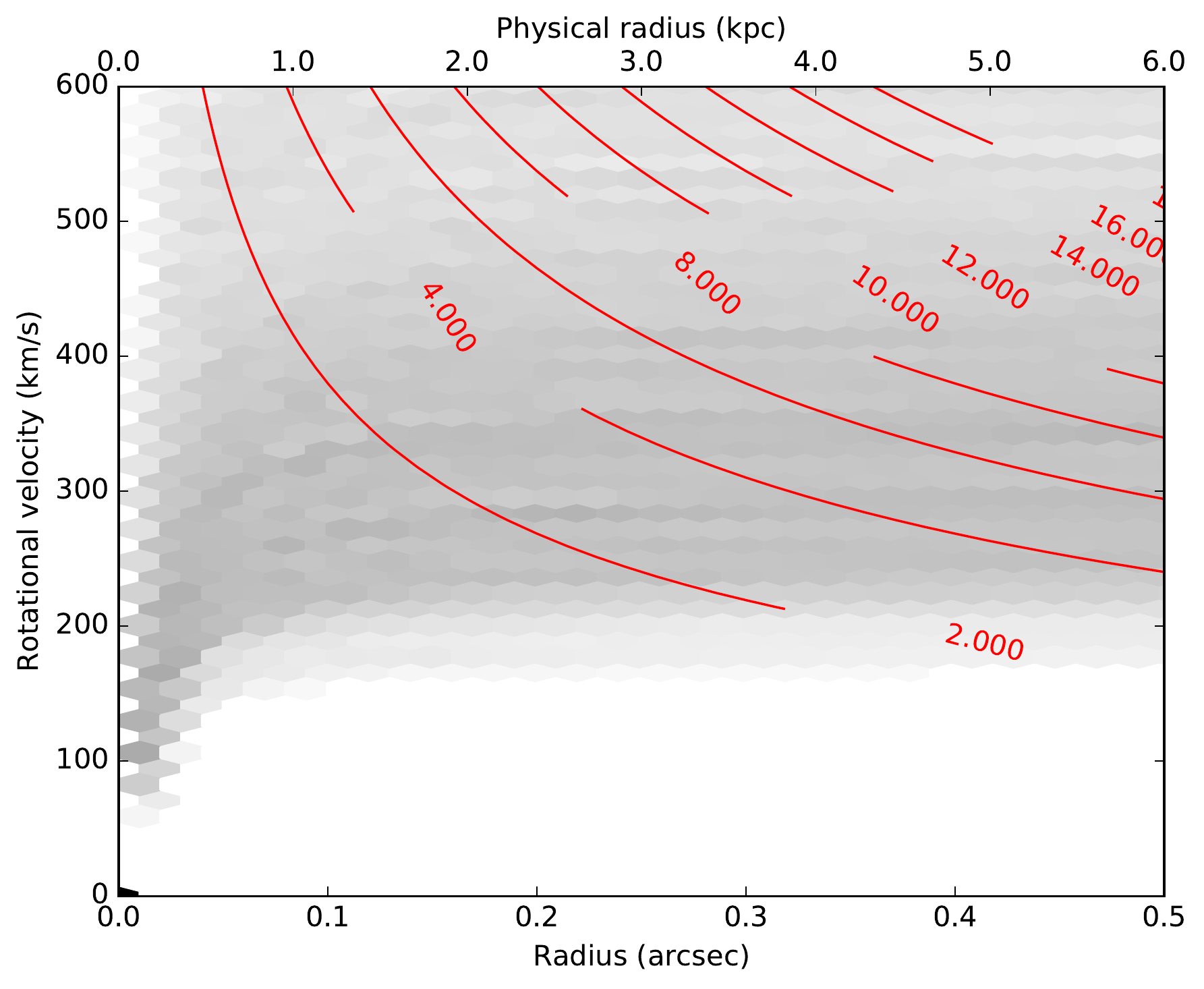}
}
\caption{Dynamical modeling results for HZ9. A single Gaussian component was adopted as continuum model. See Figure \ref{fig:dyn_mod_HZ10} for further details.}
\label{fig:dyn_mod_HZ9}
\end{figure*}

%We carry out dynamical modeling of the [C{\sc ii}] line emission from  HZ10 and HZ9 using a rotating disk model generated by \texttt{KinMS} \citep{KinMS}, fitted to the visibility data using \texttt{GALARIO} \cite{GALARIO} and \texttt{Multinest} \citep{Multinest},  using the method\footnote{ Python code available at https://github.com/pavesiriccardo/UVmodeldisk} previously described by \cite{CRLE}. We model the continuum as two and one Gaussian components for HZ10 and HZ9, respectively.
We simultaneously fit a rotating disk model generated by KinMS \citep{KinMS} to the line emission and a simple continuum model (one and two Gaussian components for HZ9 and HZ10, respectively). 
We model the line emission intensity as a Gaussian profile and the rotation curve as a ``tangent" function parametrised by the maximum velocity and the half-maximum radius (Table~\ref{table_1comp_dyn}). We fit a total of 18 parameters for HZ10 (also including line flux, disk center along each coordinate and continuum sizes, fluxes and position for both components) and 12 parameters for HZ9 (also including line flux, disk center along each coordinate and continuum size and flux) as afforded by the available signal-to-noise ratio. The data, median parameter model (indistinguishable from the best-fit model) and residuals are shown in Figures~\ref{fig:dyn_mod_HZ10} and \ref{fig:dyn_mod_HZ9}, together with the derived probabilistic constraints to the rotation curve as a function of radius and the implied dynamical masses enclosed within that radius.

Because of the limitation of assuming a single disk model, we note that substantial uncertainties regarding the detailed dynamics of HZ10 affect our inference as evidenced by the non-negligible residual structure after model fitting. 
We use the disk model scale-length and rotation curve to derive dynamical mass estimates within the half-light radius, by adopting the measured rotational velocity. We do not apply corrections for velocity dispersion because the physical origin of the apparent dispersion is uncertain (particularly in the case of HZ10, for which two distinct components may be partly responsible for the line broadening). We estimate that these systematic corrections may be as large as $\sim50\%$, toward increasing the dynamical masses inferred by fitting a rotating disk, based on the measured gas dispersion ($\sigma\sim90\pm10\,$km s$^{-1}$ and $220\pm10\,$km s$^{-1}$ for HZ9 and HZ10, respectively). We obtain $\sim(6.1\pm0.7)\times10^{10}\,M_\odot$ for HZ10 and only an approximate estimate of  $5^{+5}_{-3}\times10^{10}\,M_\odot$ for HZ9, within the half light radius of the [C{\sc ii}] emission. Our results agree within the uncertainties with previous estimates based on tilted-ring modeling in the image plane by \cite{Jones17}, although our uncertainty estimates are significantly more conservative due to the larger number of fitted parameters.

\begin{table}[htb]
\centering{
\caption[]{Results of dynamical modeling for our sample galaxies.}
\label{table_1comp_dyn}
%\resizebox{.5\columnwidth}{!}{%
\begin{tabular}{c|ccc|ccc}
\hline 
&&HZ10&&&HZ9&\\
Parameter (Units)&16th perc.&50th perc.&84th perc.&16th perc.&50th perc.&84th perc.  \\
\hline \noalign {\smallskip}
Gas dispersion (km s$^{-1}$)&210&218&226&80&89&100\\
Emission FWHM (arcsec)&0.84&0.88&0.91&0.47&0.51&0.55\\
Maximum velocity (km s$^{-1}$)&380&430&510&300&445&750\\
Velocity scale length (arcsec)&0.12&0.18&0.28&0.014& 0.03& 0.08\\
Inclination (degrees)&59&61&63&13&21&31\\
Position angle (degrees)&$-7$&$-5$&$-3$&71&77&83\\
\hline \noalign {\smallskip}
\end{tabular}
%\tablecomments{$^a$: The maximum velocity and velocity scale parameters are highly correlated because the rotation curve is poorly constrained.}
}
\end{table}

\clearpage
\bibliographystyle{aasjournal}

\bibliography{HZ10_biblio}

\end{document}